\documentclass[a4paper]{article}

\usepackage[utf8]{inputenc}   
\usepackage[T1]{fontenc}      
\usepackage[english]{babel}  
\usepackage{amsmath}
\usepackage{parskip}
\usepackage{amssymb}
\usepackage{multicol}
\usepackage{hyperref}
\hypersetup{
    colorlinks,
    citecolor=blue,
    filecolor=black,
    linkcolor=black,
    urlcolor=black
}
\usepackage{lscape}

\renewcommand{\thefootnote}{\fnsymbol{footnote}}

\usepackage[squaren,Gray]{SIunits}
\usepackage{floatrow}

\usepackage[a4paper]{geometry}

\usepackage{parskip}
\usepackage[numbers,sort,compress]{natbib}

\newcommand\blfootnote[1]{%
  \begingroup
  \renewcommand\thefootnote{}\footnote{#1}%
  \addtocounter{footnote}{-1}%
  \endgroup
}
                              
\usepackage{graphicx,wrapfig}
\usepackage[font=footnotesize,labelfont=bf]{caption}
\usepackage{float}  

\begin{document}                            

\title{Skinny emulsions take on granular matter} 
\author{Ana{\"i}s Giustiniani\textit{$^{a}$}, Simon Weis\textit{$^{b}$}, Christophe Poulard\textit{$^{a}$}, Paul H. Kamm\textit{$^{c}$},\\ Francisco Garc{\'i}a-Moreno\textit{$^{c}$}, Matthias Schr\"oter\textit{$^{d}$}, Wiebke Drenckhan\textit{$^{e*}$}
\blfootnote{Corresponding author: wiebke.drenckhan@ics-cnrs.unistra.fr}}

\date{}
\maketitle


\sloppy   


\begin{abstract}
Our understanding of the structural features of foams and emulsions has advanced significantly over the last 20 years. However, with a search for "super-stable" liquid dispersions, foam and emulsion science employs increasingly complex formulations which create solid-like visco-elastic layers at the bubble/drop surfaces. These lead to elastic, adhesive and frictional forces between bubbles/drops, impacting strongly how they pack and deform against each other, asking for an adaptation of the currently available structural description. The possibility to modify systematically the interfacial properties makes these dispersions ideal systems for the exploration of soft granular materials with complex interactions.

We present here a first systematic analysis of the structural features of such a system using a model silicone emulsion containing millimetre-sized polyethylene glycol drops (PEG). Solid-like drop surfaces are obtained by polymeric cross-linking reactions at the PEG-silicone interface. Using a novel droplet-micromanipulator, we highlight the presence of elastic, adhesive and frictional interactions between two drops. We then provide for the first time a full tomographic analysis of the structural features of these emulsions. An in-depth analysis of the angle of repose, local volume fraction distributions, pair correlation functions and the drop deformations for different skin formulations allow us to put in evidence the striking difference with "ordinary" emulsions having fluid-like drop surfaces. While strong analogies with frictional hard-sphere systems can be drawn, these systems display a set of unique features due to the high deformability of the drops which await systematic exploration.
\end{abstract}
\vspace{1cm}



\renewcommand*\rmdefault{bch}\normalfont\upshape
\rmfamily
\vspace{-1cm}

\footnotetext{\textit{$^a$ Laboratoire de Physique des Solides, CNRS, Univ. Paris-Sud, Universit\'e Paris-Saclay, Orsay Cedex 91405, France}}

\footnotetext{\textit{$^b$ Institut f\"ur Theoretische Physik I, Friedrich-Alexander-Universit\"at, 91058 Erlangen, Germany}}

\footnotetext{\textit{$^c$ Helmholtz-Centre Berlin for Materials and Energy GmbH, Berlin, Germany}}

\footnotetext{\textit{$^d$ Institute for Multiscale Simulation, Friedrich-Alexander-Universit\"at, 91052 Erlangen, Germany.}}

\footnotetext{\textit{$^e$ Institut Charles Sadron, CNRS UPR22 - Universit\'e de Strasbourg, Strasbourg, France}}

\section{Introduction}
 
Liquid dispersions consist of discrete gas bubbles or liquid drops which are tightly packed within a continuous liquid phase. For simplicity and coherence with our later discussion, we shall call the bubbles/drops "soft grains" from now on. Coalescence between these soft grains is avoided (or at least reduced) by the addition of interfacially active agents (small molecular weight surfactants, polymers, proteins, etc.). These create a protective monolayer on the grain surface which, additionally, reduces the interfacial tension  $\gamma$ down to a stabiliser-dependent value $\gamma$. In many systems, $\gamma$ can be assumed constant to a first approximation and is of the order of $\gamma \approx (1-100) \times {10}^{-3}$ N/m\cite{eastoe2000dynamic}. Characteristic grain sizes of interest here are in the range of $100-1000 \; \mu$m. The associated characteristic interfacial energies $E = \gamma S$ (where S is the surface area of the grain) are more than 10 orders of magnitude higher than thermal energies $kT$. Characteristic density differences are $100-1000$ kg/m$^{3}$, which implies that also the potential energies of these grains in the carrier fluid are at least nine orders of magnitude higher than thermal energies. Ensembles of these soft grains are therefore a-thermal, out-of-equilibium systems, implying all the complexity associated with the physical description of these kind of systems as we know it from granular media physics \cite{Song2008,van2009jamming,torquato2010jammed,andreotti:13,behringer2015jamming,bi2015statistical,jaeger:15,baule2016edwards,luding2016granular}. Foams and emulsions with sufficiently large bubbles/drops are therefore often  described as "soft granular media" \cite{weaire2007foam,van2009jamming,katgert2013jamming}.

As in hard granular media, the key parameter for the description of foams and emulsions is the global volume fraction $\Phi_\text{g}$ of the dispersed phase. Granular packings, composed of monodisperse hard spherical particles with frictional interactions\cite{schroeter:17}, are mechanically stable for values of $\Phi_\text{g}$ between approximately 0.55 \cite{onoda:90,Jerkins2008,Song2008,farrell:10,silbert:10} and 0.64 \cite{Scott1969,anikeenko_polytetrahedral_2007,torquato:00,jin_first-order_2010,kapfer_jammed_2012,francois:13,baranau_random_close_2014,rietz:18}. The lower boundary is often referred to as Random Loose Packing (RLP), the upper as Random Close Packing (RCP). In contrast, frictionless particles such as drops or bubbles do not form mechanically stable packings below $\Phi_\text{g}$ = 0.64 due to the absence of tangential forces. This lower boundary is commonly referred to as the \textit{isostatic point}, \textit{jamming transition} (assuming an increasing $\Phi_\text{g}$), or \textit{rigidity loss transition} (assuming a decreasing $\Phi_\text{g}$). However, due to their compressibility, soft grains can obtain volume fractions up to $\Phi_\text{g} \approx 1$ at which point the grains are fully polyhedral.  

Most foams and emulsions which have been considered in the past \cite{weaire1999physics,cantat2013foams} are stabilised by interfacially active agents which create fluid-like grain surfaces with constant interfacial tension and negligible solid friction/adhesion. The structural features of these kind of systems close to jamming and at very high volume fractions may now be considered well understood \cite{drenckhan2015structure,weaire1999physics,cantat2013foams}. For example, the jamming transition occurs when grains have on average $Z = 6$ neighbours, which is the minimum contact number necessary to fix all degrees of freedom. For spherical, disordered, monodisperse grains this contact number is reached at $\Phi_\text{g} \approx 0.64$. Nevertheless, a certain number of subtle questions remain open concerning the influence of the non-locality of the interaction potentials between bubbles/drops \cite{hohler2017many,weaire2017bubble} and the description of intermediate volume fractions \cite{drenckhan2015structure}.

A number of recent developments in the search for super-stable foams and emulsions \cite{rio2014unusually,salonen2014interfacial,heim2015stabilisation,heim2014rupture} has led to an increased use of stabilising strategies which create solid-like grain surfaces. Such interfaces are obtained in numerous ways, either by using specifically designed agents (certain particles, proteins or polymers) or by creating gels of agents at the interface via chemical or physical cross-linking\cite{Giustiniani2016}. The resulting "skin-like" interfaces have a finite, solid-like interfacial elasticity (i.e. an applied strain induces a static interfacial stress). They also lead to solid friction between the soft grains, and in some cases, they render the grains adhesive. The resulting normal and tangential forces lead to complex grain interactions and additional mechanical constraints which influence strongly how the soft grains pack and deform around and beyond jamming. The influence of such interfacial forces between large ($R>10-100$ $\micro$m) bubbles or drops has been only sporadically treated in the literature. For example, the influence of cohesive forces (adhesion) was investigated by Bruji\'c \textit{et al.} \cite{Feng2013,Pontani2012,Pontani2013,Pontani2016,Jorjadze2011} and Hadorn \textit{et al.} \cite{Hadorn2012}. Bruji\'c \textit{et al.} \cite{Pontani2016} reported that drops having adhesive patches at their interface reached mechanical stability at volume fractions of $\Phi_\text{g} \approx$ 0.55, while their non-adhesive counterparts exhibited a volume fraction of $\Phi_\text{g} \approx$ 0.64, as expected. In another study \cite{Jorjadze2011}, they used silicone oil-in-water emulsions stabilised by a charged surfactant (SDS) whose concentration was varied to induce and control attractive depletion forces $\langle F_\text{d} \rangle$ between the drops. They showed that by varying $\langle F_\text{d} \rangle$, they were able to vary the global volume fraction from $\Phi_\text{g} \approx$ 0.74 down to $\Phi_\text{g} \approx$ 0.6.

It is therefore important to start systematic investigations into the structural features which characterise the packings of (very) soft grains with such complex interactions. Additional importance of such investigations arises from the necessity to avail of granular systems in which the "hardness" of the grains, as well as the friction and adhesion between them, can be controlled explicitly (and ideally independently) in order to explore systematically the relationship between the grain interactions and the obtained grain packings. Important advances have been established in the case of hard spheres in recent years, both  in computer simulations \cite{papanikolaou:13,liu:15,chen:16,mughal2012dense} and in experimental investigations \cite{scheel:08,lois:08,butt:09,gogelein:10,koos:12,rieser:15,hemmerle:16}.

In order to advance with the experimental investigations of soft grains, we use here a model system which consists of monodisperse polyethylene glycol drops (PEG) which are dispersed in liquid polydimethylsiloxane (PDMS). The drops are stabilised by chemical cross-linking of the drop surfaces, which creates a solid-like skin. Here we only provide a short summary of the key features of the skin formation (Section \ref{SkinProperties}) in order to concentrate on the analysis of the influence of this skin on the drop packing. More information on the skin formation has been provided in Giustiniani \textit{et al.} \cite{Giustiniani2016}. In Section \ref{DropInteraction} we investigate the tangential and normal drop-drop interactions using a simple two-drop model experiment. We then establish a link between the drop interactions and the structural properties of the obtained emulsions using X-ray tomography (Section \ref{PackingAnalysis}). We show that the droplet surfaces have an elastic skin which leads to non-negligible friction and adhesion between drops - the degree of which we vary by the addition of dodecane in the PDMS. We put in evidence that the structural features of these "skinny emulsions" are very different to those obtained in "ordinary", frictionless emulsions. However, while many of the observed features resemble those of packings of frictional, hard grains (lack of spontaneous ordering, finite angle of repose, low and height-invariant volume fractions - Section \ref{subsec:Results}), certain features carry clear signatures associated with the deformability of the grains (flat radial distribution function) which ask for more systematic investigation in the future.


\section{Interfacial properties of the drops}\label{SkinProperties}

\begin{figure}[ht]
\centering
\includegraphics[width=8cm]{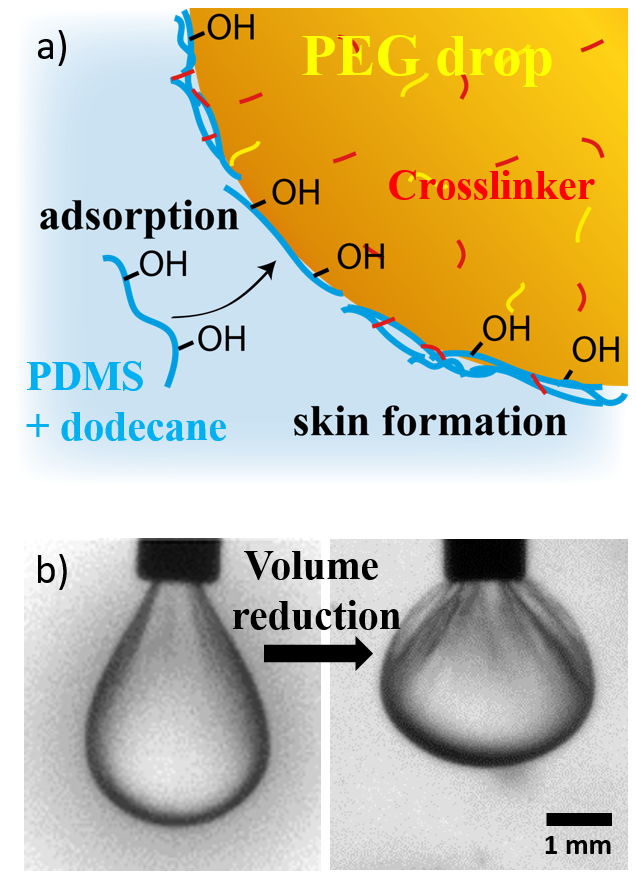}
  \caption{a) Scheme of the composition of the PEG-in-silicone emulsions stabilised by a reactive stabilisation approach. b) Image of a PEG drop containing the crosslinker/catalyst created in Sylgard 184 base and D$_4$ with 10\% of dodecane before and after aspiration of some of the liquid of the drop.}\label{fig:Stabilisation}
\end{figure}

Here we study drops of PEG with molecular weight $M_\text{w}$ = 400 g/mol (with viscosity $\eta$ = 0.96 Pa$\cdot$s and density $\rho$ = 1.128 g/mL) inside a liquid PDMS (Sylgard 184{\textregistered} base) in which we add a small quantity of octamethylcyclotetrasiloxane, also called D$_4$ (non-reactive equivalent of the Sylgard 184{\textregistered} curing agent). For more detailed information on the formulation see Section 6.1. The stabilisation of the PEG drops is made possible by the creation of a skin-like interface which arises from chemical reactions at the drop surface in the presence of a catalyst. For this purpose we add the crosslinker/catalyst molecule (Platinum(0)-1,3-divinyl-1,1,3,3-tetramethyldisiloxane complex solution 0.1 M in poly(dimethylsiloxane), vinyl terminated) to the PEG drop from which it diffuses to the drop surfaces. There it provokes two types of reactions (Figure \ref{fig:Stabilisation}a). First, the C-OH end groups of the PEG and Si-OH groups present in the Sylgard 184 can react, creating PDMS-b-PEG copolymers at the interface while releasing water. The Si-OH bonds can also react with each other, leading to a cross-linking reaction which creates a solid-like PDMS "skin" at the interface. We established the link between these reactions and the emulsion stability for a model system in Giustiniani \textit{et al.} \cite{Giustiniani2016}. This "skin" renders the emulsions indefinitely stable and can be visualised by the presence of wrinkles at the surface of a PEG drop after reduction of the drop volume (Figure \ref{fig:Stabilisation}b). The addition of dodecane (at 5$\%$ and 10$\%$ in weight) to the PDMS phase allows us to modify the skin properties and thus the interactions between the drops, which we study in Section \ref{DropInteraction}. The presence of dodecane also modifies the viscosity which varies between 0.52 Pa$\cdot$s for 10$\%$ of dodecane and 1.91 Pa$\cdot$s for 0$\%$ of dodecane (see Table \ref{Table:SamplesEmulsions} and Section \ref{MaterialsAndMethods} for more information).

\section{Interactions between two drops}\label{DropInteraction}

\begin{figure*}[ht!]
\centering
\includegraphics[width=17cm]{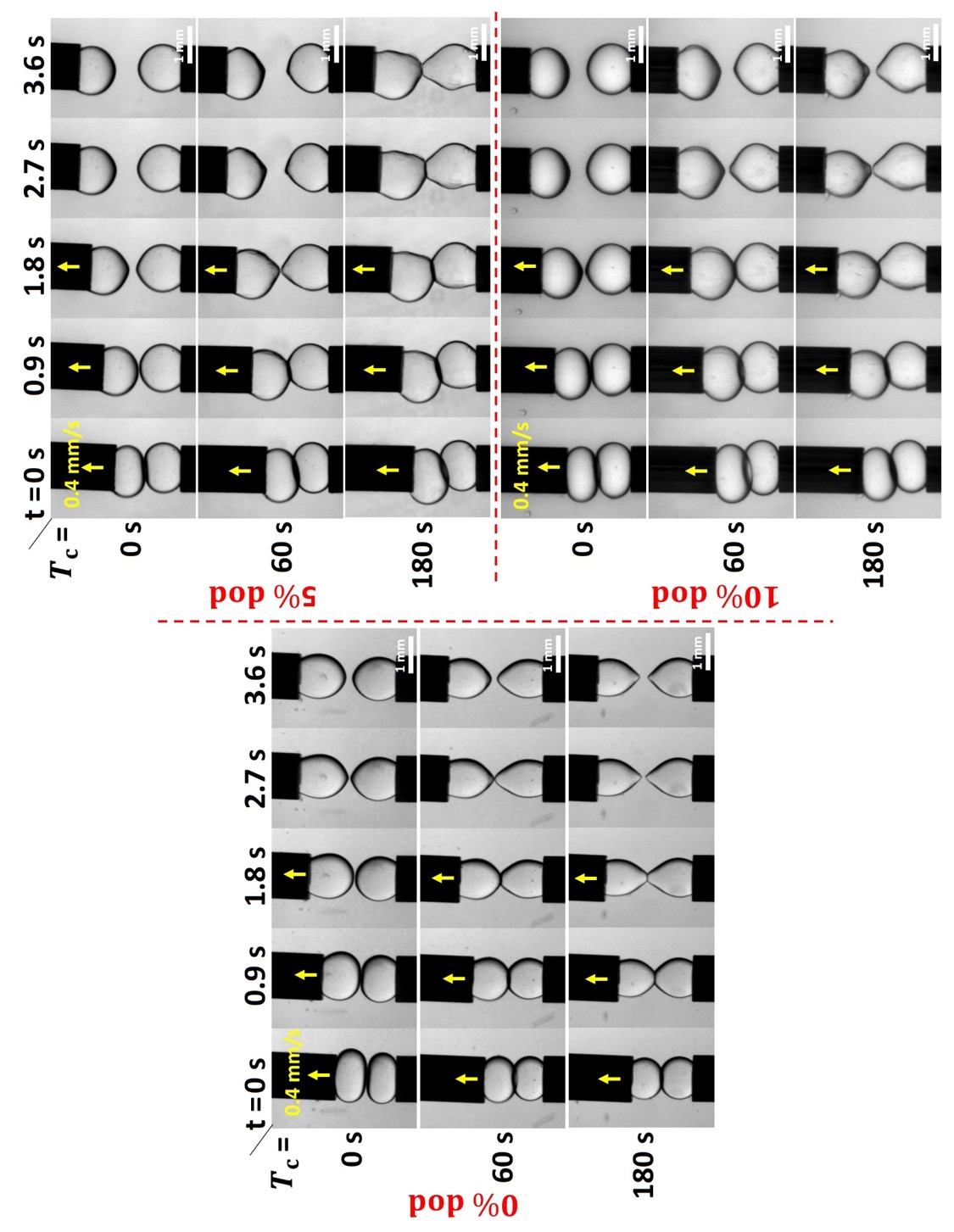}
  \caption{Image sequences of the separation of the drops for a contact time of $T_\text{c}$ = 0, 60 and 180 s and with 0\%, 5\% and 10\% of dodecane in the continuous phase, at $v$ = 0.4 mm/s after a stabilisation time of $T_s$ = 120 s.}
  \label{fig:Adhesion-v_04}
\end{figure*}

We study the interaction between two drops using a home-built, fully automated "double drop experiment" (DDE, Figure \ref{fig:Adhesion-v_04}) which allows to generate two drops of controlled volume and to move them with respect to each other in all three spatial directions at controlled speed and chosen moments. Using the DDE we are able to simulate the interactions which arise between two drops during the emulsion generation while avoiding the impact of the presence of other drops. To do so, two syringes are mounted on a holder and connected at the top to a motor which delivers or retrieves a controlled volume of the liquid in the syringe. On the other side, the syringes are connected to two needles, one straight and one curved. The syringe attached to the curved needle can move vertically and horizontally, which allows in a first step to fix its position exactly in line of the other syringe. All experiments presented here are made at room temperature. The experiment is piloted by a \textit{Labview} program, which allows to fix the desired volume of each drop (here systematically chosen at $V$ = 2 $\micro$L), and the approach speed $v$ of the two needles.

With this setup, we conducted two types of experiments: the detection of adhesive properties of the interfaces and of interfacial friction. Both experiments start with a stabilisation phase of time $T_\text{s}$ = 120 s at a distance $d_\text{max}$ between the two needle tops during which the drops remain separated and the interfaces are undergoing the chemical reactions described in Section \ref{SkinProperties}. This time was chosen since it represents roughly the time it takes for the droplets to arrive on the emulsion during the generation (Section \ref{PackingAnalysis}). This phase is followed by a contact phase of time $T_\text{c}$ during which the drops are put in contact at a distance $d_\text{min}$ = 1.5 mm between the two needle tops, which is smaller than twice the drop diameter and leads to the creation of a flat contact zone between the drops. Once the drops have been in contact during the time $T_\text{c}$, we can investigate the adhesive properties of the drop surfaces by separating them at controlled speed (Section \ref{Sec:Adhesion}). To study the friction between the interfaces, we slide them horizontally against each other and look at the final drop shape after relaxation (Section \ref{Sec:Friction}).

\subsection{Characterisation of the adhesion between two drops}\label{Sec:Adhesion} 
\label{sec:adhesion}

In order to observe the signature of adhesive forces between two drops we leave them in contact for a time $T_\text{c}$ before pulling them apart from each other at a traction velocity of $v_\text{t}$ = 0.4 mm/s. We repeat this experiment for three contact times ($T_\text{c}$ = 0, 60 and 180 s) and for three dodecane concentrations (0, 5 or 10\% dodecane). The contact times are chosen after approximation of the characteristic initial contact times during emulsion generation depending on the viscosity and density of the continous phase. Image sequences of the two-drop adhesion experiments are shown in Figure \ref{fig:Adhesion-v_04}. We can see that for 0\% of dodecane in the continuous phase (Figure \ref{fig:Adhesion-v_04} left), the drops are elongated before they relax slowly due to strong viscous forces arising from the high viscosity of the silicone phase. The surface of the drops remains smooth throughout the drop detachment. This is not the case for higher dodecane concentrations (Figure \ref{fig:Adhesion-v_04} right). First of all, we see that especially after long contact times, the two drop surfaces seem to stick to each other. This does not only lead to a noticeable deformation of the drops during separation, but also to the formation of wrinkles on the drop surfaces which are a clear sign of the presence of a solid-like skin. Moreover, even though the wrinkles disappear after drop detachment, the drop shape remains strongly non-Laplacian, being again indicative of a solid-like interface. All effects seem strongest for a 5\% dodecane. This is due to competing effects resulting from the simultaneous influence of the dodecane concentration on the bulk viscosity and on the skin formation, as discussed in Section \ref{sec:DiscussionDoubleDrop}. Proper quantification of the adhesive forces - for example through the measurement of contact angles between drops - is not yet possible with our set-up due to the difficulty of ensuring perfect alignment of the two drops. Nevertheless, one can conclude from the image sequences presented in Figure \ref{fig:Adhesion-v_04} that non-negligible adhesive forces exist between the drops which increase with dodecane concentration and with contact time. 
 
It should be noted that in this experiment, the separation of the drops is the result of strong external driving forces. Separation forces in the emulsion are weaker, which is why such adhesive forces may lead to permanently stuck drops inside the emulsion, as will be discussed in Section \ref{PackingAnalysis}.

\begin{figure*}[ht]
\centering
\includegraphics[width=17cm]{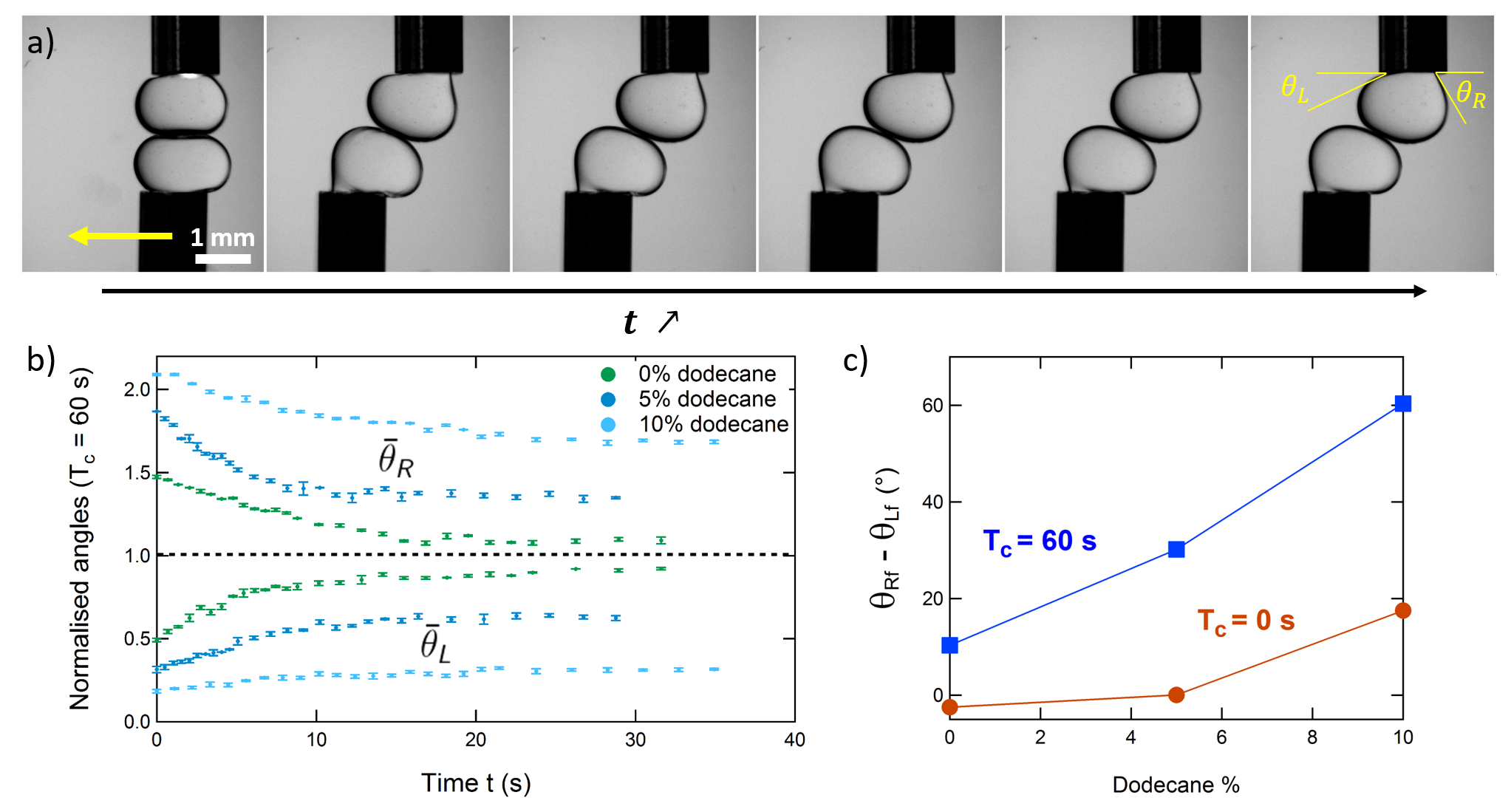}
  \caption{a) Image sequence of the interfacial friction experiment: the drops are put in contact for $T_\text{c}$ = 60 s, then the top drop is moved rapidly towards the right and slides over the bottom drop. A qualitative measure of the interfacial friction is obtained from the time evolution of the left $\theta_\text{L}$ and right $\theta_\text{R}$ angle between the top drop and the needle as shown in the image on the right. b) Evolution of the normalised angle $\bar{\theta}$ (c.f. eq.~\ref{eq:theta_norm}) with time for $T_\text{c}$ = 60 s. c) Difference between the final values of $\theta_\text{R}$ and $\theta_\text{L}$, $\theta_\text{R,f}-\theta_\text{L,f}$, depending on the percentage of dodecane in the continuous phase for two contact times ($T_\text{c}$ = 0 s and $T_\text{c}$ = 60 s).}
  \label{fig:Friction}
\end{figure*}

\subsection{Characterisation of the tangential forces between the drops}\label{Sec:Friction}
\label{sec:friction}

For these experiments, two drops are created in the continuous phase with either 0, 5 or 10\% of dodecane. At the end of the stabilisation time $T_\text{s} = 120$ s, they are brought in contact during $T_\text{c}$ = 0 or 60 s. Then, the top needle is moved to the right by a distance $\Delta X$ = 0.7 mm at a speed $v$ = 0.1 mm/s. The result of this needle motion is that the drops slide over each other. Figure \ref{fig:Friction}a shows a resulting image sequence for the example of 5$\%$ dodecane. Once the needle stops moving, we quantify the relaxation towards the equilibrium position of the drop on the top by measuring the left $\theta_\text{L}$ and right $\theta_\text{R}$ angles between the drop and the needle as shown in the far right picture in Figure \ref{fig:Friction}a. An example of the measurement of $\theta_\text{L}(t)$ and $\theta_\text{R}(t)$ is given in the Supplementary Materials (Figure \ref{fig:Angles-fit}), for 0\% of dodecane in the continuous phase and $T_\text{c}$ = 60 s. The $\theta(t)$ curves can be fitted by an exponential function $\theta (t)=\theta_f+\theta_0\exp(-t/\tau)$, which allows to extract the final value $\theta_f$ and the relaxation time $\tau$ towards $\theta_\text{f}$, for both $\theta_\text{L}$ and $\theta_\text{R}$. We did not find any measurable influence neither of the dodecane percentage nor of the contact time $T_\text{c}$ on the relaxation time $\tau$, which is of the order of 10 s. This may be due to competing effects between the viscosity of the bulk phase and the skin properties. For direct comparison, we plot in Figure \ref{fig:Friction}b the time evolution of the angles normalised by the mean of their final values for $T_\text{c}$ = 60 s:
\begin{equation}
\label{eq:theta_norm}
\bar{\theta}(t)=\frac{\theta_\text{L,R}(t)}{\langle \theta_\text{f} \rangle}\text{,}
\end{equation}
where $\langle \theta_\text{f} \rangle=\frac{\theta_\text{L,f}+\theta_\text{R,f}}{2}$. We see that even though the evolution of $\theta_\text{L}$ and $\theta_\text{R}$ exhibit a similar behaviour in time for all three concentrations, their final state differ.

Figure \ref{fig:Friction}c shows the evolution of the difference between the final angles $\theta_\text{R,f}-\theta_\text{L,f}$ with the concentration of dodecane and for two different contact times $T_\text{c}$ = 0 and 60 s. We can see that this difference in angle increases both with the dodecane concentration and with the contact time. For example, for $T_\text{c}$ = 0 s, $\theta_\text{R,f}-\theta_\text{L,f}$ is close to zero for 0 and 5\% of dodecane, and increases to $\theta_\text{R,f}-\theta_\text{L,f}\approx$ 20$^{\circ}$ for 10\% of dodecane. All observations indicate that the mechanisms at the origin of the angle difference are stronger for higher dodecane concentrations.

\begin{figure}[t]
\centering
\includegraphics[width=9cm]{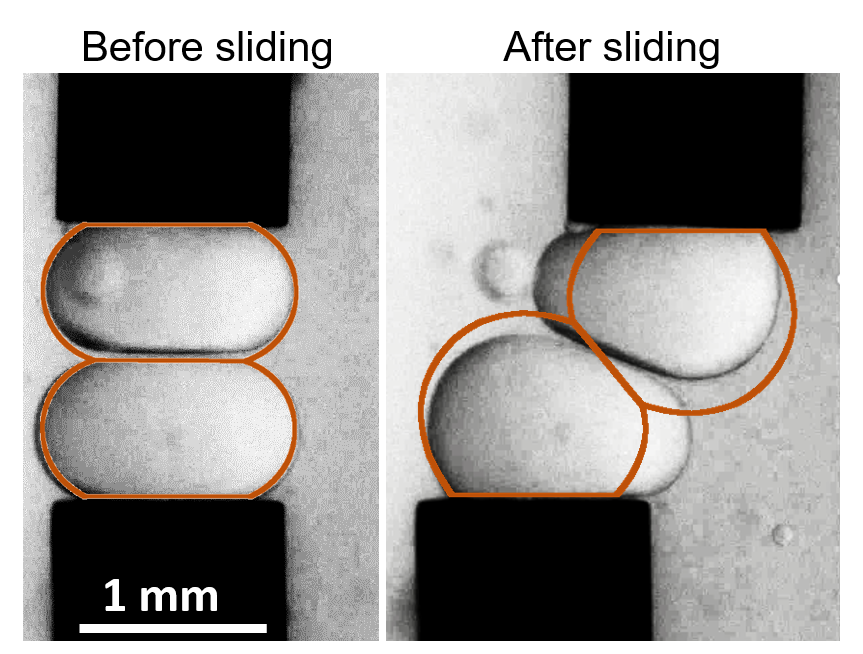}
  \caption{Superimposition of experimental images and shape profile of simulations using Surface Evolver (red lines) of the sliding of drops, for 5\% of dodecane in the continuous phase. The simulations assumes that the drop shapes are only controlled by surface tension forces.}
  \label{fig:Numerics_comparison}
\end{figure}

In Figure \ref{fig:Numerics_comparison}, we compare the images of the two drops before and after a sliding experiment (5$\%$ dodecane, $T_\text{c}$ = 60 s) with simulations (superimposed in red). These simulations are done using the program Surface Evolver\footnote{available at \url{http://facstaff.susqu.edu/brakke/evolver/evolver.html}}. Surface Evolver is used for the modeling of liquid surfaces shaped by various forces and constraints at equilibrium. It searches for the configuration of a given system with the minimal energy. We used the simulations to predict the equilibrium shapes which the two drops would have if only surface tension governs the shape of the drops, i.e. no surface elasticity, no adhesive forces and no friction at the interface. The simulations therefore predict the shapes one would expect for  "ordinary" emulsion drops with fluid-like surfaces. This comparison shows that forces other than surface tension are at play in the relaxation of the top drop towards its equilibrium position. These forces are not measurable in the drop shapes before sliding, but appear clearly after the sliding experiment. 

\subsection{Discussion}
\label{sec:DiscussionDoubleDrop}

Sections \ref{Sec:Adhesion} and \ref{Sec:Friction} confirm that the drop surfaces grow a solid-like skin \cite{Giustiniani2016} whose formation seems to be accelerated by the presence of dodecane. Moreover, we see that in contrast to "ordinary" emulsions, our drops display the clear presence of normal attractive and tangential forces. These forces increase with the presence of dodecane and with the contact time of the drops. 

An important question here is whether we can talk separately about adhesion and/or friction at the interface. Experiments in Section \ref{Sec:Adhesion} unambiguously show normal attractive forces, i.e. adhesive processes between the drops. Now we need to determine whether we can say that the phenomena observed in Section \ref{Sec:Friction} are frictional forces or not. Indeed, adhesion could also prevent the sliding of the drops over each other. Soft granular materials with lower volume fraction were indeed observed in the presence of adhesive forces between drops in an emulsion with no friction between the drops \cite{Jorjadze2011}. For hard granular systems, friction is added between the spheres by surface roughening of the beads \cite{pohlman:06,Jerkins2008,farrell:10,utermann:11,back:11,sheng:16,gillemot:17}, while adhesion forces are kept equal to zero. 

In our case, the forces are induced by the presence of the polymeric interface and their magnitude seems to vary by the presence of dodecane in the continuous phase. In polymeric materials, either polymer melts or elastomers (crosslinked network of polymer chains), frictional and adhesive processes can both arise from entanglement and chemical or physical bonding of the polymer chains at the interface. This means that in our case, we cannot truly separate the contributions of adhesion and friction\cite{maeda2002adhesion}. Indeed, the friction between two materials depends on many parameters \cite{Myshkin2006}: surface roughness, surface chemistry, lubrication, environment, temperature, etc. In our case, the interface is not roughened at the interface scale, though it is at the molecular scale. The lubrication is important in our system, where it is provided by the continuous phase inbetween the two drops. The viscosity of the continuous phase is thus important, and should certainly play a non-negligible role in the differences observed between the different dodecane concentrations in the experiment where two drops slide over each other for a given contact time. The dodecane also changes the surface chemistry of the interface by swelling the polymeric skin. This might make it easier for polymers to entangle at the interface when the drops are in contact, which would explain why the adhesion between the drops depends on the dodecane concentration in the continuous phase.

For a given concentration of dodecane, the mechanisms responsible for the interactions between the drops seem to be dependent on the contact time. A first hypothesis is given by the fact that the polymeric gels at the drop surfaces penetrate each other with time. As the contact time increases, entanglement is made easier between the polymers of each drop. Additionally, since the reactions at the interface between the drops are not stopped after the stabilisation time, an alternative hypothesis would be that the reactions occur between the skin of two touching drops, "connecting" them chemically. Finally, as the addition of a solvent in the continuous phase changes its density and viscosity, the reaction rates might change at the interface. Indeed, the reaction rate depends on the diffusion coefficient of the molecules involved, which, in a liquid, is inversely proportional to the viscosity. Lowering the viscosity of the continuous phase increases the reaction rate at the interface and thus the number of covalent bonds between two drops for a given contact time.

Last but not least, the decreased viscosity of the PDMS and different solvent conditions which lead to faster reaction kinetic would also lead to a faster growth of the skin's elasticity, which adds a non-negligible resistance to drop deformation (as can be seen in Figure \ref{fig:Adhesion-v_04}). Hence, in order to fully interpret our observations, one needs to take into account: surface tension, surface elasticity, adhesion and friction. This is a truly complex exercise which will be tackled in future work. Nevertheless, the experiments conducted in this section provide us with a clear proof that all these contributions exist and that they increase with dodecane concentration. This is why the dodecane concentration will remain one of the key control parameter in the following sections.

\begin{figure}[t]
\centering
\includegraphics[width=10cm]{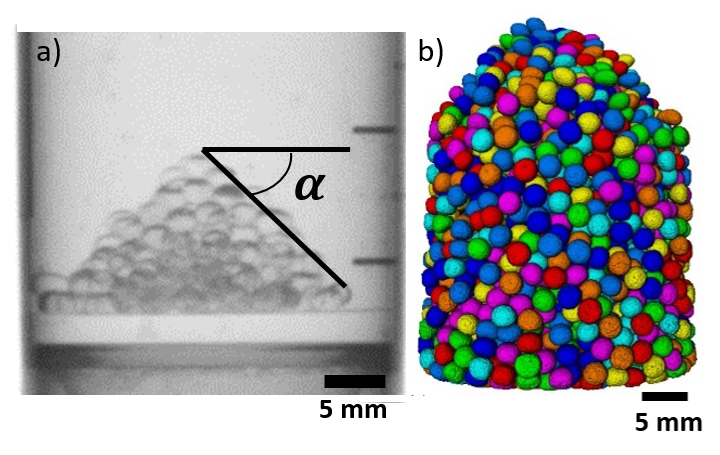}
  \caption{a) Photograph of the angle of repose of a PEG-in-silicone emulsion with 5\% of dodecane in the continuous phase. b) 3D volume rendering of a PEG-in-silicone emulsion with 5\% of dodecane in the continuous phase obtained from X-ray tomography.} \label{fig:AngleOfRepose_tomo}
\end{figure}

The complex drop interactions have a great impact on the packing of the drops. A first indication is shown by the clear presence of a finite angle of repose $\alpha$ which is created by sedimenting drops during the emulsion generation (described in Section \ref{MaterialsAndMethods}). A photograph of an example with 5$\%$ dodecane is shown in Figure \ref{fig:AngleOfRepose_tomo}a. Table \ref{Table:AngleOfRepose} provides the values of the angle of repose for the different dodecane concentrations, measured as indicated in Figure \ref{fig:AngleOfRepose_tomo}a. These results indicate that the concentration of dodecane does play a measurable role in the macroscale properties of the packing of drops. We notice that $\alpha$ is low but non negligible for 0\% of dodecane in the continuous phase, while $\alpha$ increases greatly with the addition of dodecane from $\approx$ 10$^{\circ}$ for 0\% of dodecane up to $\approx$ 40$^{\circ}$ for 5 and 10\% of dodecane. For a rigorous interpretation of these results one needs to take into account conflicting effects of the dodecane concentration. On the one hand, the viscosity of the continuous phase decreases with increasing dodecane concentration. Hence the drops arrive faster on the packing and may therefore have less time to rearrange before the arrival of a new drop. The associated decrease of the contact times between drops may reduce the importance of frictional/adhesive interactions (as shown in Figures \ref{fig:Adhesion-v_04} and \ref{fig:Friction}c) and hence, potentially, the angle of repose. However, since the skin interactions increase quite significantly with dodecane concentration, we believe that this effect outweighs the effect of the contact time. For example, as can be seen in Figure \ref{fig:Friction}c, tangential interactions between two drops at 10\% dodecane concentration after a contact time of $T_\text{c}$ = 0 s are more important than the tangential interactions of 0\% dodecane after a contact time of $T_\text{c}$ = 60 s.

The observation of a very low but finite $\alpha$ = 4.6$^{\circ}$ $\pm$ 1$^{\circ}$ was found experimentally for wet foams\cite{lespiat2011jamming}. Numerical simulations also found a low value of $\alpha$ of 5.76$^{\circ}$ $\pm$ 1$^{\circ}$ on systems of nearly rigid beads with no static friction\cite{peyneau2008frictionless}. Taboada \textit{et al.}\cite{taboada2006additive} used numerical simulations with polydisperse disks with varying friction coefficient to show that $\alpha$ consistently fell between 4$^{\circ}$ and 7$^{\circ}$ for frictionless disk, while it increased significantly with the friction coefficient of the disks. Pohlman \textit{et al.}\cite{pohlman:06} also found both experimentally and computationally that the angle of repose of a packing of rigid beads increased with increasing friction coefficient. This shows again the influence of adhesive and/or tangential forces on the organisation of drops within an emulsion which will be discussed in the following.

\begin{table}[t]
\begin{center}
\caption{Angle of repose measured for each dodecane percentage $\%Dod$ during the generation of the emulsions, given with the statistical error.} \label{Table:AngleOfRepose} 
    \begin{tabular}{ | c | c | }
    \hline
    $\%Dod$ & Angle of repose $\alpha$ ($^{\circ}$) \\ 
    \hline
    \hline     
    0\% &  10.8 $\pm$ 0.5 \\ 
    \hline
    5\% &  40.8 $\pm$ 2.1 \\ 
    \hline
    10\% &  43.4 $\pm$ 2.3 \\ 
    \hline     
    \end{tabular}
\end{center}
\end{table}

\section{Packing of frictional and adhesive emulsion drops}\label{PackingAnalysis}

\subsection{Sample generation and characteristics}\label{Samples}

Here, we study the packing of monodisperse PEG drops in a continuous phase composed of Sylgard 184{\textregistered} base, D$_4$ and dodecane (5 or 10\% in weight). The emulsions are generated at room temperature by letting the PEG drops settle one by one through the PDMS phase (Section \ref{MaterialsAndMethods}). Depending on the dodecane concentration, the sedimentation time of the drops is between 100 and 300 s, which explains the value of the stabilisation time $T_s$ = 120 s chosen in Section \ref{SkinProperties}. Interestingly, the emulsions generated with $0\%$ dodecane were not stable, indicating that the protective skin around the drops was not formed sufficiently rapidly in order to avoid drop coalescence.

\begin{figure}[t]
\centering
\includegraphics[width=8cm]{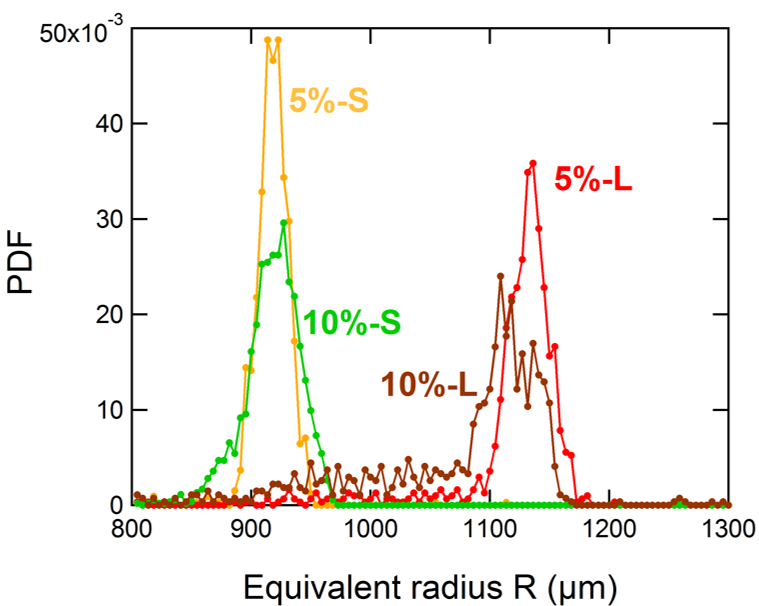}
  \caption{Distribution of the equivalent radius of the drops in the different emulsions.} \label{fig:size-distribution}
\end{figure}

We use two different drop sizes for each dodecane concentration. The average equivalent radius $\langle R \rangle$ and the associated standard deviation $\sigma_R$ for each sample is calculated from the tomography results (Section \ref{XRaytomo}) and is given in Table \ref{Table:SamplesEmulsions}. Both are obtained from the distribution of the equivalent radius\footnote{The equivalent radius consists in calculating the radius $R$ of the drop from its volume $V$ which is known thanks to X-ray tomography by the relation $V=\frac{4}{3}\pi R^3$. I.e. the drop is assumed to be spherical for the size calculations.} (Figure \ref{fig:size-distribution}). Table \ref{Table:SamplesEmulsions} also lists the polydispersity index $PI=\sigma_R/\langle R \rangle$ of the drops, the amount of dodecane added in the continuous phase $\%Dod$ (in weight \%) and the viscosity $\eta$ of the continuous phase. For simplicity, in the rest of the study we refer to the drop size as either $S$ (small, $R \approx$ 910 $\micro$m) or $L$ (large, $R \approx$ 1100 $\micro$m).

Figure \ref{fig:size-distribution} shows the distributions of the equivalent radius $R$ for all four samples (Section \ref{Samples} and Table \ref{Table:SamplesEmulsions}). The samples 5\%-S, 5\%-L and 10\%-S have a low polydispersity index $PI$. However, the sample 10\%-L has a higher $PI$ due to the presence of smaller drops in the emulsions, which might have been created during the generation. The results given for this sample are therefore to be interpreted cautiously.

\begin{table*}[t]
\centering
\caption{Summary of the main parameters of the emulsions studied in Section \ref{PackingAnalysis}, with \%Dod the dodecane percentage in the continuous phase, $\eta$ the viscosity of the continuous phase, $R$ the equivalent radius of the drops, $PI$ the polydispersity index of the drop sizes in the emulsions, $\Phi_\text{g}$ the global volume fraction and $\sigma_{\Phi_\text{l}}$ the width of the distributions of the local volume fraction $\Phi_\text{l}$ obtained by fitting the distributions with Gaussian functions, and $\langle \beta_0^{20} \rangle$ the average value and $\sigma_{\beta}$ the width of the distributions of the parameter $\beta_0^{20}$ characterising the drop deformations.} \label{Table:SamplesEmulsions} 
    \begin{tabular}{ | c | c | c | c | c | c | c | c | c | c |}
    \hline
    Sample name & $\%Dod$ & $\eta$ (Pa$\cdot$s) & $\langle R \rangle$ ($\micro$m) & $\sigma_R$ ($\micro$m) & $PI$ & $ \Phi_\text{g} $ & $\sigma(\Phi_\text{l})$ & $\langle \beta_0^{20} \rangle$ & $\sigma_{\beta}$ \\ 
    \hline
    \hline
     
    5\%-S & 5\% & 0.83 & 915.0 & 19.1 & 2.1\% & 0.48 & 0.08 & 0.68 & 0.16 \\ 
    \hline
    5\%-L & 5\% & 0.83 & 1121.0 & 75.3 & 6.7\% & 0.50 & 0.10 & 0.60  & 0.17 \\ 
    \hline
    10\%-S & 10\% & 0.52 & 904.5 & 63.3 & 7.0\% & 0.34 & 0.10 & 0.57 & 0.25 \\ 
    \hline
    10\%-L & 10\% & 0.52 & 1071.1 & 118.4 & 11.1\% & 0.40 & 0.16 & 0.39 & 0.22 \\ 
    \hline    
    \end{tabular}
\end{table*}

\begin{figure}[t]
\centering
\includegraphics[width=9cm]{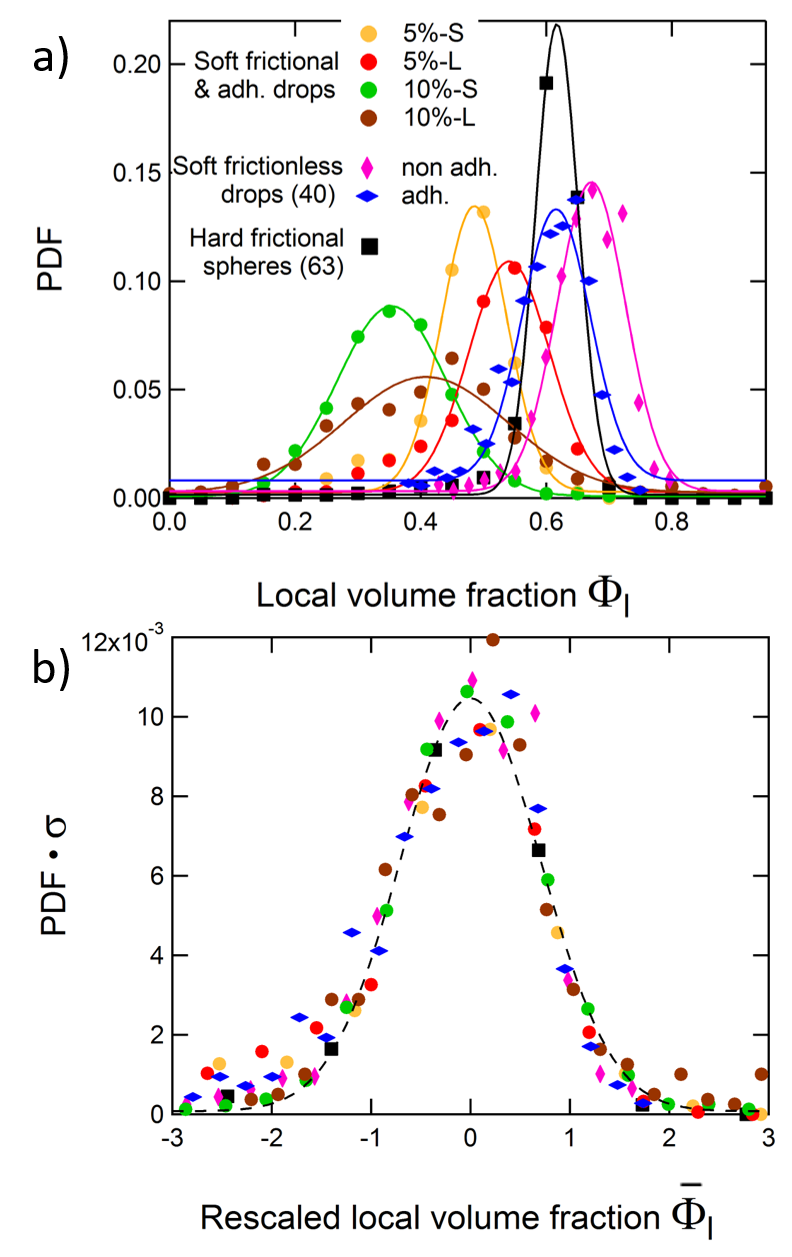}
  \caption{a) Distribution of the local volume fraction $\Phi_\text{l}$ of our emulsions compared with the distributions obtained for frictionless emulsion drops\cite{Jorjadze2011} presenting adhesive interactions or not ($F_\text{d}$ = 0 pN and $F_\text{d}$ = 27 pN respectively) and hard frictional spheres\cite{Weis2017xray}, fitted with a Gaussian function. b) Rescaled distribution of the local volume fraction $\bar{\Phi_\text{l}}$ in a) (using Equation (\ref{PhiResc})).} \label{fig:phi_local}
\end{figure}

\subsection{3D-structural analysis of the emulsions}\label{subsec:Results}

In the following, we study in detail the structural properties of the emulsions. To do this, we used a lab-based X-ray tomography setup (Section \ref{XRaytomo}), which allows to obtain 3D reconstructions (Figure \ref{fig:AngleOfRepose_tomo}b) of the samples listed in Table \ref{Table:SamplesEmulsions}. We determine the distribution of the local volume fraction $\Phi_\text{l}$ and the global volume fraction $\Phi_\text{g}$ of the emulsions, and analyse in detail the organisation and deformation of the drops within the emulsions. 

The \textbf{local volume fraction $\Phi_\text{l}$} of the emulsions is defined as the ratio between the volume of a given drop and the total volume belonging to that drop. The former is directly obtained  from the X-ray analysis, the latter is computed using a \textit{Set-Voronoi tessellation} \cite{Schaller2013SetVoronoi}. The Set-Voronoi tessellation is a generalisation of the Voronoi tessellation which is applicable to particles of arbitrary shape. It is based on the surface-to-surface distances of the drops, instead of the center-to-center distances as in the commonly used \textit{Point Voronoi tessellations}. The tessellation is performed using the program \textit{Pomelo}\footnote{Software available at \url{http://theorie1.physik.uni-erlangen.de/research/pomelo/}} \cite{weis:17} based on the voxelized surface of the drops, which was calculated from the tomography data.

Figure \ref{fig:phi_local}a shows the distributions of the local volume fraction $\Phi_\text{l}$ obtained for the four emulsions presented in Table \ref{Table:SamplesEmulsions} together with distributions of local volume fractions obtained for frictionless emulsion drops (with or without adhesion) \cite{Jorjadze2011} and for hard frictional spheres\cite{Weis2017xray}. We see that they can all be fitted by Gaussian distributions, though in the case of the samples 5\%-S and 5\%-L we can see an overpopulation of loosely packed drops (tail-like structure) on the left side of the distribution. These low  $\Phi_\text{l}$ tails mean that the 5\% packings have some interesting geometrical feature, which we have not identified yet. The distribution of the 10\%-L sample is noisy - probably due to the large $PI$ - but can still be described with a Gaussian function. 

The \textbf{global volume fraction $\Phi_\text{g}$} of the whole packing is the harmonic mean of the local volume fractions of all particles in the sample\cite{weis:17}. $\Phi_\text{g}$  and the standard deviation $\sigma({\Phi_\text{l}})$ of the local volume fraction distributions are listed in Table \ref{Table:SamplesEmulsions}. The value of $\Phi_\text{g}$ decreases with increasing dodecane percentage for a given drop radius, being an indication of a looser packing.

In order to compare these distributions independently of the value of $\Phi_\text{g}$, we consider the rescaled local volume fraction 
\begin{equation}\label{PhiResc}
\bar{\Phi_\text{l}} = \frac{\Phi_\text{l} - \Phi_\text{g} }{\sigma({\Phi_\text{l} })}\text{.}
\end{equation}

The result is shown in Figure \ref{fig:phi_local}b. We can see that within the data scatter all distributions collapse on one master curve (black dashed line). This means that the rescaled distribution of the local volume fraction of our system is not distinguishable from a packing of hard frictional spheres or soft frictionless drops, whether they show adhesive surfaces or not. This representation of the local volume fraction thus seems to be independent of the interactions between the spheres and of the global volume fraction of the packings within the error bars of our experiments.

\begin{figure}[t]
\centering
\includegraphics[width=11cm]{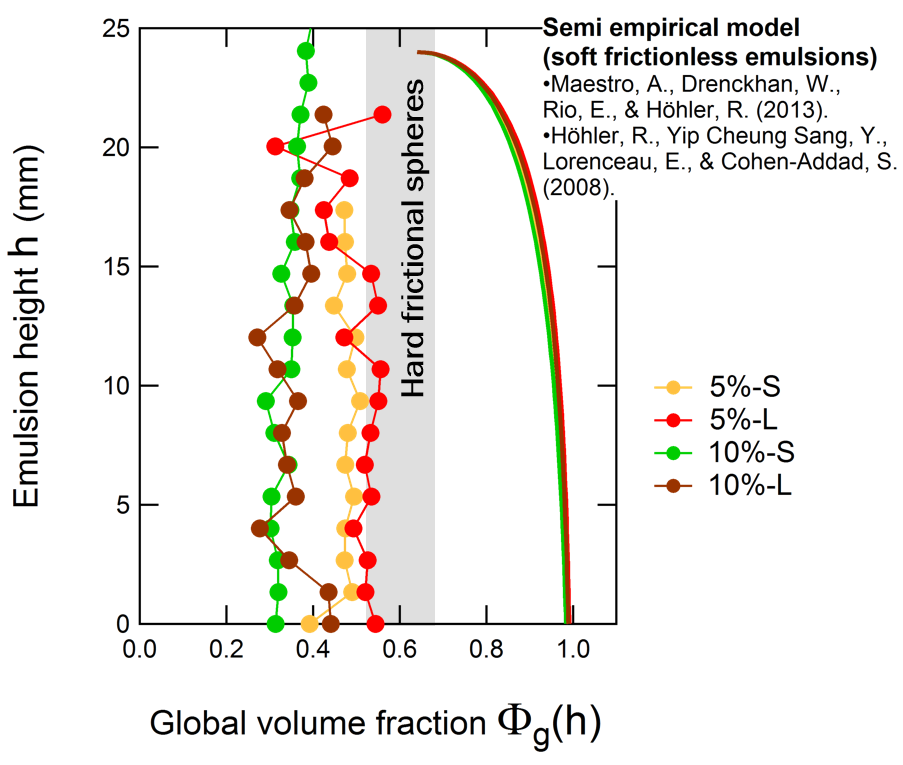}
  \caption{Evolution of the global volume fraction $\Phi_\text{g}(h)$ with the emulsion height $h$ measured for the emulsions presented in Table \ref{Table:SamplesEmulsions}. The semi empirical model for soft, frictionless emulsion drops developed by Maestro \textit{et al.} \cite{Maestro2013} and H\"ohler \textit{et al.} \cite{Hohler2008} is also shown, along with the range of values of the global volume fraction expected for hard frictional spheres (in gray).}
  \label{fig:phi_h}
\end{figure}

Based on the local volume fraction and the position of a droplet, the global packing fraction can also be calculated within different bins at an emulsion height $h$, where $h=0$ corresponds to the bottom of the emulsion. We thus obtain the variation of the global volume fraction with height $\Phi_\text{g}(h)$.
Figure \ref{fig:phi_h} shows $\Phi_\text{g}(h)$ for the samples presented in Table \ref{Table:SamplesEmulsions}. We can clearly see that $\Phi_\text{g}(h)$ does not follow the semi-empirical model provided by Maestro \textit{et al.} \cite{Maestro2013} (following a study by H\"ohler \textit{et al.} \cite{Hohler2008}) for surfactant-stabilised, frictionless emulsions with the same drop size, density difference and interfacial tension (measured\cite{giustiniani_thesis} to be $\gamma$ = 6 mN/m). On the contrary, the profiles are constant with $h$. For their model, Maestro \textit{et al.} took into account only surface tension effects and neglected any surface elasticity of the drops. In our system, however, the surface elasticity of the drops is not negligible and will resist strongly the drop deformations (Figure \ref{fig:Stabilisation}b). The surface elasticity may be taken into account by considering an effective surface stress, which modifies the osmotic pressure of the emulsion and hence it's resistance to deformation under gravity. Investigations into such matter have been performed in the past on Pickering emulsions covered by elastic nano-particle layers \cite{arditty2003some}. However, following this line of argument, a constant volume fraction profile of an emulsion in contact with a pool of continuous phase implies that drops are undeformed. Yet, as we shall see later, the drops in our system are quite strongly deformed and cannot be approximated by non-deformable spheres. The constant volume fraction of soft spheres may instead be explained by the Janssen effect\cite{janssen:95,sperl:06,vanel:99,Bertho2003,wambaugh:10,back:11} which states that in the presence of friction, the contact forces between the spheres redirect the weight, i.e. the pressure, towards the walls of the container, and therefore the pressure in the bulk is independent of the height. Since in Section \ref{SkinProperties} we put in evidence the presence of a non-negligible friction and adhesion, we therefore assume that the elasticity of the drops together with the tangential and normal attractive forces between them act simultaneously to explain the constant volume fraction $\Phi_\text{g}(h)$ with the emulsion height $h$. For such a hypothesis we need to make the assumption that friction (and potentially adhesion) is also present between the drops and the container wall. While we have not measured this explicitly, the presence of friction between the drops and the wall is reasonable since both are rough at the microscopic scale. Also, viscoelastic materials, such as the polymeric skin around the drops, tend to stick to rigid surfaces like the container wall\cite{creton2003pressure}. The magnitude of either contribution is, however, unknown to us at this stage and should be characterised in future work.

Additionally to the invariance of the volume fraction with height, we also find low values of $\Phi_\text{g}$ compared to the lowest values of the global fraction $\Phi_\text{RLP}\approx$ 0.55 known for hard spheres with friction. A slight underestimation of the volume fraction may come from the fact that we do not know precisely the thickness of the skin around the drops, i.e. we do not know their effective size. However, closer analysis of the thickness of the flat "films" between neighbouring drops in the tomography images allow us to estimate the thickness of the skin to be < 50 $\mu$m. This would add a correction of 0.08 to the values of $\Phi_\text{g}$ in Figure \ref{fig:phi_h}, which can therefore not explain the very low values observed for 10\% of dodecane. Values of $\Phi_\text{g}$ below the loose-packing density of hard, frictional spheres were observed by Liu \textit{et al.} \cite{Liu2011} for spheres connected by liquid bridges, i.e. in the presence of adhesive forces between the spheres. Very low values of $\Phi_\text{g}$ are usually observed for colloidal systems with attractive interactions between particles (flocculated)\cite{arditty2003some,scales1998shear,sonntag1987elastic}. For example, Arditty \textit{et al.} \cite{arditty2003some} studied oil-in-water emulsions stabilised by solid particles, which showed values of $\Phi_\text{g}$ down to $\Phi_\text{g} \approx$ 0.2, attributed to the observed (and relatively controlled) flocculation of the droplets. As we saw in Section \ref{DropInteraction}, the presence of the polymeric skin around the droplets in the emulsions induces both normal attractive and tangential forces between the drops in the presence of dodecane. The low values of $\Phi_\text{g}$ are in that sense not incoherent.

Figure \ref{fig:phi_h} also seems to show that the amount of dodecane has an impact on the global volume fraction of drops in the emulsion. Indeed, the values of $\Phi_\text{g}$ for 10\% of dodecane are lower than the values for 5\% of dodecane, with a difference of $\Phi_\text{g}(5\%)-\Phi_\text{g}(10\%)\approx$ 0.15 for a given drop radius. The study of the drop-drop interactions in Section \ref{DropInteraction} showed that their magnitude depends on $\%Dod$, and studies in the literature show that the global volume fraction decreases with the magnitude of the sphere-sphere interaction \cite{Jorjadze2011}. The decrease of $\Phi_\text{g}$ with $\%Dod$ in our case may thus be explained by the increasing interaction between the drops. The width of the distributions of $\Phi_\text{l}$ also increases with the strength of the adhesion between emulsion droplets in the study of Jorjadze \textit{et al.} \cite{Jorjadze2011}, which correlates well with our results.

\begin{figure}[ht]
\centering
\includegraphics[width=10cm]{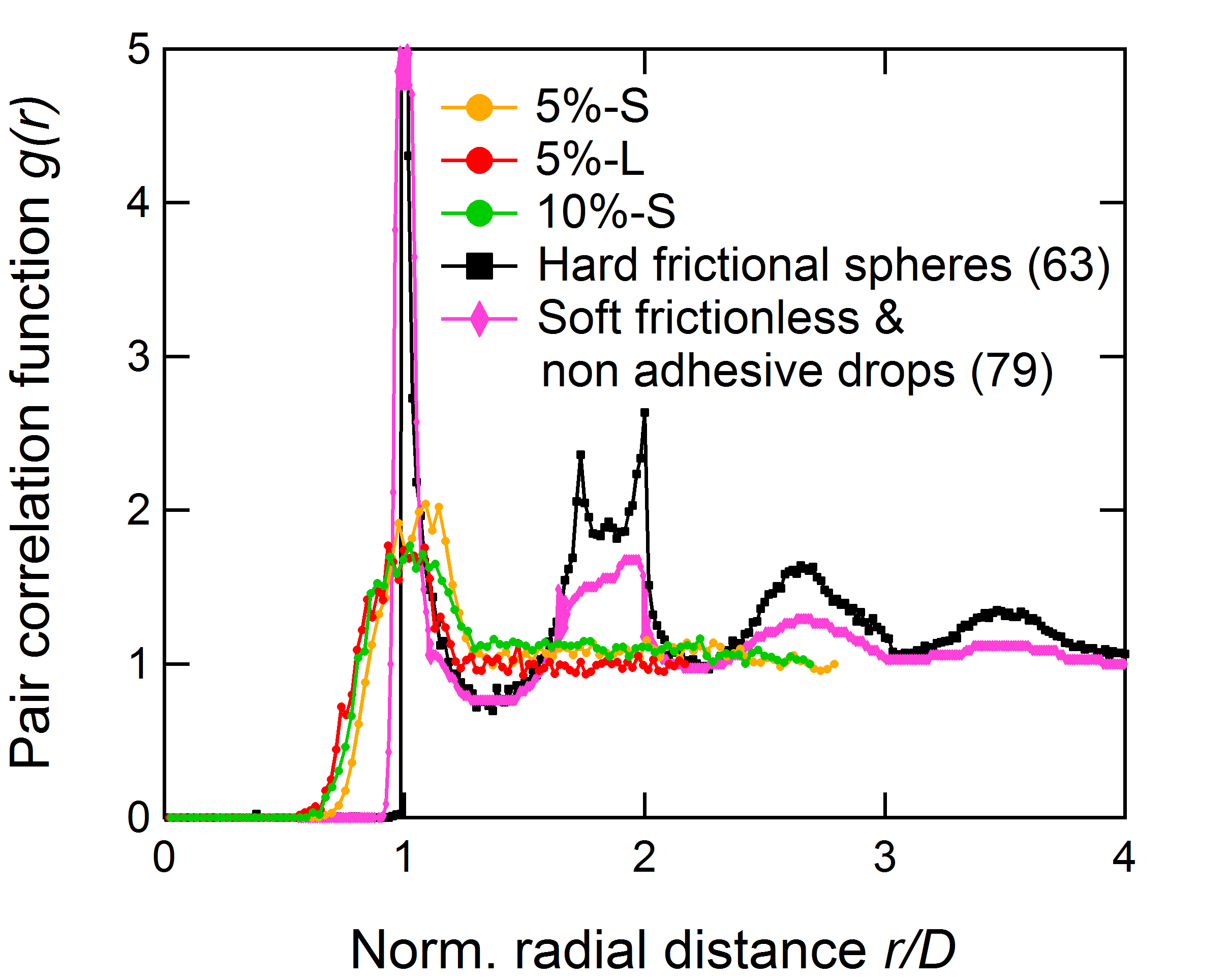}
  \caption{Pair correlation function $g(r)$ of our emulsions, compared with the $g(r)$ obtained for both frictionless and non adhesive emulsion drops \cite{zhang2005jamming} and hard frictional spheres\cite{Weis2017xray}.} \label{fig:g_r}
\end{figure}

We probed the local fluctuations in density in the emulsions by calculating the \textbf{pair correlation function $g(r)$}, which is also often called the radial distribution function. Figure \ref{fig:g_r} shows the $g(r)$ for our emulsions, compared with the $g(r)$ obtained for a packing of hard frictional spheres, using the same data presented in Figure \ref{fig:phi_local}b\cite{Weis2017xray}, and by Zhang \textit{et al.} \cite{zhang2005jamming} from simulations for a frictionless emulsion. In the case of the frictionless drops and the frictional hard spheres, we can clearly observe the expected sharp peak at $r/r_0$ = 1 and the two peaks at $r/r_0$ = $\sqrt{3}$ and $r/r_0$ = 2, which are characteristic of amorphous packings of monodisperse spheres. 

In the case of our emulsion drops we make two important observations. Firstly, the peak at $r/r_0=1$ is broad, with a width $\sigma_{g}\approx$ 0.45 which is independent of the dodecane concentration $\%Dod$. This cannot be explained by the polydispersity of the drops since - as shown in Table \ref{Table:SamplesEmulsions} - the PI are low for the samples represented in Figure \ref{fig:g_r}. However we can notice that the peak at $r/r_0=1$ for the soft, frictionless and non adhesive emulsions from \cite{zhang2005jamming} is broader than the one for the hard frictional spheres, which is due to the deformability of the emulsion drops. The broad peak calculated for our emulsions could hence be the result of the deformation of the drops. 
Secondly, we also notice the complete absence of any further characteristic peak in the $g(r)$ of our emulsions, which indicates no correlations in the positional order of the drops with respect to each other in the packing. This flat $g(r)$ may also be explained by the deformation of the drops, which is why we now proceed to a close analysis of the drop shapes.

To analyse the deformation of the drops, the anisotropy of the drops is calculated.
Anisotropies in the shape of objects can be characterised using Minkovski Tensors \cite{Schroder2011}. 
Note that Minkowski Tensors are normally used to measure the anisotropy of the packing (e.g. of the Set-Voronoi cells of a packing). In constrast we use Minkowski Tensors here to analyse the shape of the particles itself. The Minkowski Tensors are a generalisation of scalar-valued Minkowski functionals \cite{Gerd:2009_mink,Kapfer_2011}, which are an established method for the description of the morphology and structure of various physical systems. The program \textit{Karambola}\footnote{Software available at \url{http://theorie1.physik.uni-erlangen.de/research/karambola/}} was used to perform these calculations for our emulsion drops.

While Minkowski functionals are not sensitive to anisotropic effects, Minkowski tensors are.
It has been shown that all six Minkowski tensors show a resemblance to the moment of inertia tensor for different mass distributions. Here we will focus on one specific Minkowski tensor, $\mathbf{W_0^{2,0}}$, which shows a resemblance (but is different) to the moment of inertia tensor of the object that is filled with a constant volume density
\begin{equation}
\mathbf{W_0^{2,0}} = \int_K \vec{r} \otimes \vec{r}  dV\text{.}
\end{equation}

The scalar term
\begin{equation}
\beta_0^{20}=\frac{|\varepsilon_\text{min}|}{|\varepsilon_\text{max}|}\in [0,1]\text{,}
\end{equation}
 measures the degree of anisotropy of the drops through the ratio of the minimal $\varepsilon_\text{min}$ to maximal $\varepsilon_\text{max}$ eigenvalue of the Minkowski tensor $W_0^{20}$. $\beta_0^{20}$ lies in a range from 0 to 1.

For a perfectly \textit{isotropic} object, all the eigenvalues $\varepsilon$ of the Minkowski tensor have the same value. 
Thus $\beta_0^{2,0} = 1$ for an isotropic object.
If the object is anisotropic, the tensor $\mathbf{W_0^{2,0}}$ will have different eigenvalues $\varepsilon$, which will result in a $\beta_0^{2,0}$ smaller than one.

\begin{figure}[t]
\centering
\includegraphics[width=10cm]{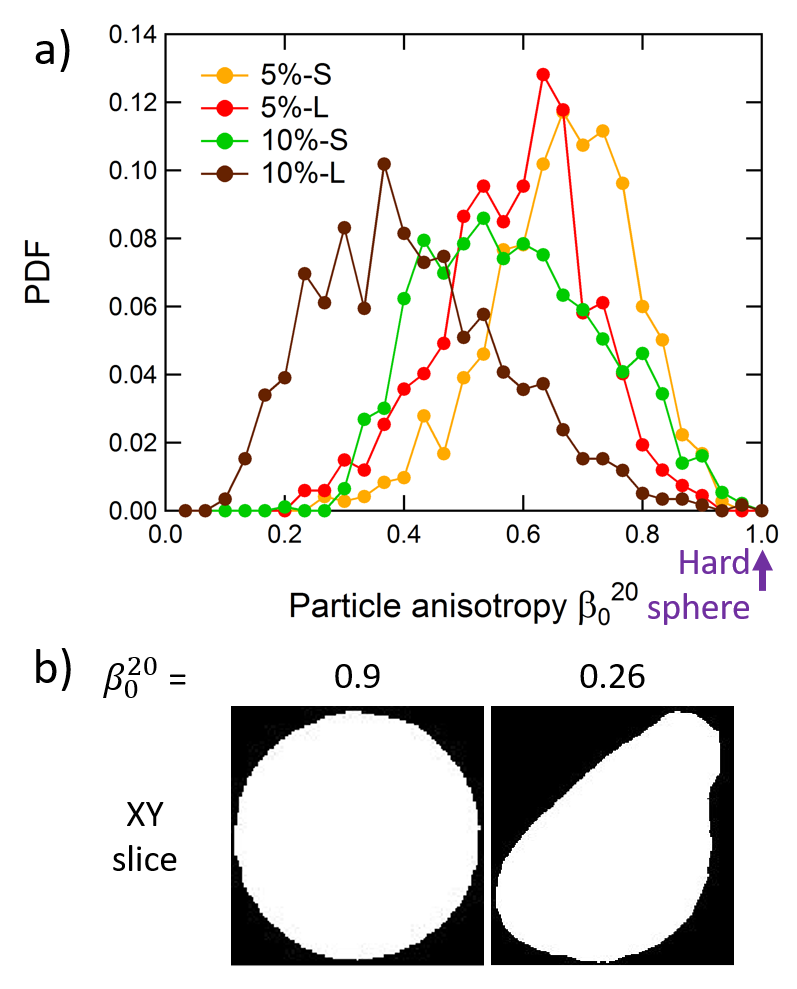}
  \caption{a) Distribution of the measured values of $\beta_0^{20}$ in the four samples. b) XY slice of drops with two extreme values of $\beta_0^{20}$ from the sample 5\%-S.} \label{fig:DropShape}
\end{figure}

The Minkowski tensor $\mathbf{W_0^{2,0}}$ and the respective anisotropy index $\beta_0^{2,0}$ are calculated for each individual droplet. 
The measure of the distributions of $\beta_0^{20}$ for the drops of the four emulsions of Table \ref{Table:SamplesEmulsions} are shown in Figure \ref{fig:DropShape}a. We can see that the distributions are indeed systematically large and centered on low values of $\beta_0^{20}$, when hard spheres systematically have $\beta_0^{20}$ = 1 (as indicated by an arrow in Figure \ref{fig:DropShape}a). Table \ref{Table:SamplesEmulsions} gives the average value $\langle \beta_0^{20} \rangle$ and the width $\sigma_{\beta}$ of each distribution grossly fitted with a Gaussian function. The amount of dodecane in the continuous phase does not seem to have a significant impact on $\langle \beta_0^{20} \rangle$. Except for the sample 10\%-L, for which the average value $\langle \beta_0^{20} \rangle$ = 0.39, the other samples exhibit similar values of $\langle \beta_0^{20} \rangle$ = O(0.6). The highly deformed drops of the sample 10\%-L are probably due to another mechanism which would also be responsible for the high polydispersity of this drop packing. Figure \ref{fig:DropShape}b shows two examples of drops with extremal values of $\beta_0^{20}$ = 0.9 and 0.26 from the sample 5\%-S. These high deformations of the drops can explain the broad peak at $r/r_0=1$ and the missing structure peaks for $r/r_0>$ 1 of the pair correlation function in Figure \ref{fig:g_r}. 

\begin{figure}[ht]
\centering
\includegraphics[width=9cm]{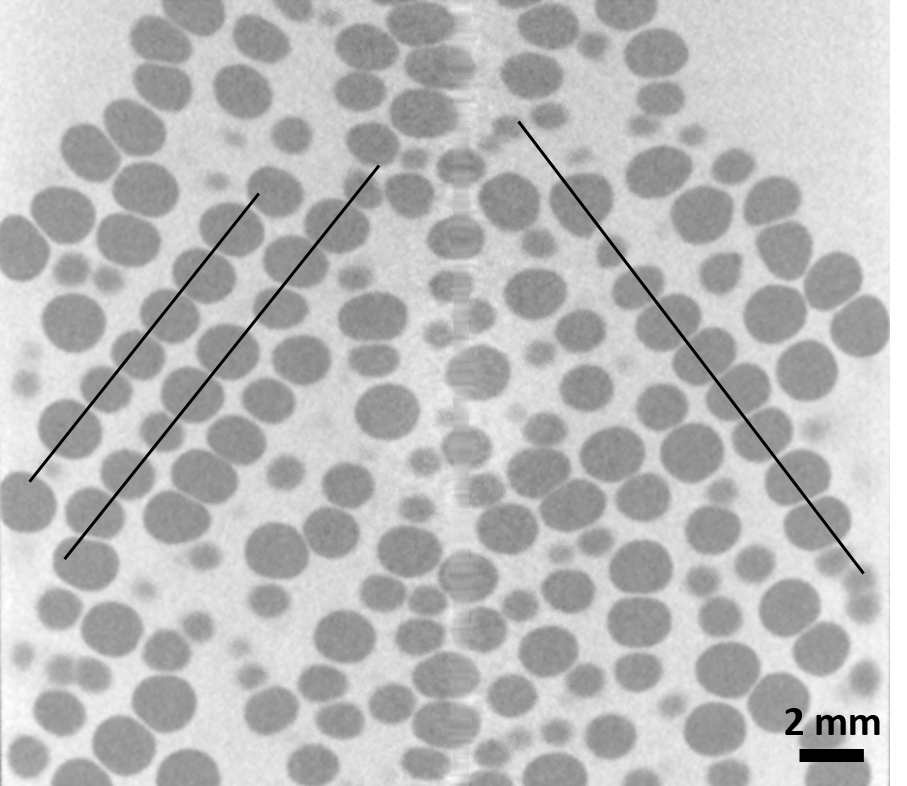}
  \caption{Vertical slice at the center of the reconstruction of the volume of the 5\%-S sample using X-ray tomography showing the presence of force chains (some of which are indicated by black lines).} \label{fig:ForceChains}
\end{figure}

To understand where such strong deformations may come from, we show in Figure \ref{fig:ForceChains} a vertical slice through the center of the reconstruction of the volume of the 5\%-S sample obtained using X-ray tomography. In this figure, we can distinguish lines of drops whose deformations follow a single direction (some of them indicated by black lines), which are similar to force chains observed in packings of hard granular spheres \cite{Anthony2005,brodu:15} or disks\cite{majmudar:05,daniels:17}. These may be sufficient to deform the drops strongly. 
This indicates that the adhesive and frictional forces between our drops might be responsible for the creation of these structures during the emulsion generation, which in turn are responsible for the low global volume fractions measured in Figure \ref{fig:phi_h}.

Finally, we were not able to measure the average number of contacts $\langle z \rangle$ between the drops because of the inability of the X-ray tomography technique to differentiate between the polymeric skin and the continuous phase.

\section{Conclusion}

The aim of this study was to present a new type of granular material, composed of soft emulsion drops surrounded by a polymeric skin. Thanks to a specially designed "double-drop" experiment (Section \ref{DropInteraction}) we could show that the skin-like interface leads to both attractive normal (adhesive) and tangential (frictional) forces between the drops. These influence directly the structure of the emulsions, for which we presented a first systematic structural analysis using X-ray tomography (Section \ref{PackingAnalysis}). This emulsion system may therefore be placed in-between the extreme cases of soft frictionless bubbles/drops and hard frictional spheres. 

The two-drop experiment allowed to investigate the influence of the dodecane on the interactions between the drops, and we were able to draw several conclusions. First, the drop separation experiment in Section \ref{Sec:Adhesion} puts in evidence adhesive forces, which seem much stronger for 5\% and 10\% of dodecane than for 0\% of dodecane in the continuous phase. Then, the drop sliding experiment in Section \ref{Sec:Friction} evidences the resistance to sliding, i.e. the presence of tangential forces between the drops. These also increase with the dodecane concentration. For this system, friction and adhesion are surely coupled in a complex manner, as discussed in Section \ref{sec:DiscussionDoubleDrop}. The presence of such complex interaction forces between the drops are very different from "ordinary" emulsion drops stabilised by low molecular weight surfactants. While much more systematic investigations need to be undertaken in the future, it seems clear that both the contact time and the amount of dodecane in the continuous phase allow to tune the magnitude of the frictional and adhesive forces between the drops for this formulation.

Having characterised the impact of the dodecane on the interaction forces between the drops, we studied in Section \ref{PackingAnalysis} the structural properties of their packings. We demonstrated a number of properties specific to this system. First of all, as for packings of frictional hard spheres, these emulsions show a finite angle of repose, which causes the final structure of the material to be impacted by the way the emulsions are generated as shown by the slope at the top of the emulsion in Figure \ref{fig:AngleOfRepose_tomo}b. The high dynamic angle of repose given in Table \ref{Table:AngleOfRepose} is a direct result of the presence of strong interactions between the drops. 

Also, the global volume fraction $\Phi_g$ of the emulsions does not vary measurably with the emulsion height, which is usually characteristic of the packing of hard spheres instead of soft emulsion drops.  We explain this by the Janssen Effect, rather than by a lack of drop deformation. Looking at the local volume fraction distribution $\Phi_\text{l}$, we were not able to find any distinguishable characteristic differences from other systems like hard frictional spheres or soft frictionless emulsion drops. However, we noticed that the global volume fraction is low compared to the volume fraction of random loose packings of frictional hard spheres ($\Phi_\text{g} <\Phi_\text{{RLP}}\approx$ 0.55). We were able to measure even lower volume fraction than reported in the literature for systems of adhesive, frictional hard spheres\cite{Liu2011}. These low volume fraction are also well explained by the strong interactions between our emulsion drops which provide the mechanical stability of the packings with a low number of contact (below isostaticity), i.e. low density of drops. The packing fraction also decreases with increasing dodecane concentration, which agrees well with the increasing strength of the drop-drop interactions shown in Section \ref{DropInteraction}. 

The analysis of the overall organisation of the drops with the calculation of the pair correlation function $g(r)$ puts in evidence a very broad first peak. No further correlation beyond the nearest neighbours have been found. In particular, we confirmed the total absence of characteristic peaks at $r/r_0$ = $\sqrt{3}$ and 2, which are normally found for random organisations of both hard frictional spheres and soft frictionless drops. We interpret both observations by the deformation of the drops. Finally, we noticed the presence of unusual alignments of drops in the emulsions, which we explain by a combination of the drop adhesion and the formation protocol of the emulsion. The flat pair correlation function and the presence of these drop alignments are thus characteristics of the packing of soft frictional and adhesive drops.

In summary, we could show that packings of soft frictional and adhesive emulsion drops resemble in many aspects those of frictional hard sphere packings. However, they also present intriguing new features, such as the flat $g(r)$ and the drop alignments, which encourage more systematic investigations.

While these experiments have provided us with first important insight, future work needs to establish approaches which provide a finer control over the skin properties of the drop surfaces. Concerning the drop-drop interactions, much more systematic experiments and computer simulations are needed to fully understand and control the complex and strongly coupled interactions which arise between the drops. It will be important to avail of systems which provide independent control of the skin elasticity, the friction and the adhesion between the drops. In this context it will be essential to understand the precise influence of the nature of the solvent used in the continuous phase. In relation to the drop organisation in the emulsions, more systematic experiments are needed in order to understand the influence of the packing protocol (frequency of arriving drops, arrival speed of drops, container shape, etc.) and various other system parameters (liquid viscosities, density differences, drop sizes, etc.). In particular, to allow comparison with a large body of available literature which reports investigations of granular systems dispersed in low viscosity media (such as air or water), it will be important to conduct future studies over a wider viscosity range.

\section*{Acknowledgements}

The authors thank Vincent Klein for the instrumentation work on the double-drop experiment, and Reinhard H\"ohler for numerous discussions related to the content of this article. We would also like to thank Myfanwy Evans and Daniel Nasato for sharing computational results to help us understand the flat $g(r)$, and Patrick K\'ekicheff for tomographic tests at the synchrotron and associated discussions. Jasna Bruji\'c is thanked for feedback on local packing fractions. We acknowledge funding from the European Research Council (ERC) under the European Union's Seventh Framework Program (FP7/2007-2013) in form of an ERC Starting Grant, agreement 307280-POMCAPS. Part of this work was produced within the IdEx Unistra framework and benefited from funding from the French National Research Agency as part of the 'Investments for the future' program.
Part of the work was funded by the German Research Foundation (DFG) through Forschergruppe FOR1548 "Geometry and Physics of Spatial Random Systems" (GPSRS) and the Cluster of Excellence Engineering of Advanced Material.

\section{Supplementary materials}
\subsection{Materials and methods}\label{MaterialsAndMethods}

The polyethylene glycol (PEG) with molecular weight $M_w$ = 400 g/mol was used as received from SIGMA-ALDRICH (CAS number: 25322-68-3). The crosslinker/catalyst (Platinum(0)-1,3-divinyl-1,1,3,3-tetramethyldisiloxane complex solution 0.1 M in poly(dimethylsiloxane),vinyl terminated) was used as received from SIGMA-ALDRICH (CAS numbers: 68478-92-2), at the concentration $C$ = 0.05 mol\% in the PEG-400. The Sylgard 184{\textregistered} is a commercial polymer supplied by DOW CORNING as a two-parts kit: a base and a curing agent. In this study, we used only the base and replaced the curing agent by octamethylcyclotetrasiloxane (also called D$_4$) (CAS number:  556-67-2) in order to maintain a similar chemistry while avoiding the solidification of the PDMS. It was used as received from SIGMA-ALDRICH, with the same proportions as usually done for the Sylgard 184{\textregistered} mix: ratio 10:1 in weight between the base and the D$_4$. We added dodecane in the continuous phase to change the drop-drop interactions (CAS number: 112-40-3), at a percentage $\%Dod$ (in weight) of the total weight of the Sylgard 184{\textregistered} base and D$_4$. All solutions are mixed at room temperature for at least 1 hour in order to ensure homogeneous mixing.

The drops are generated by dispensing the PEG/crosslinker/catalyst mixture from a syringe at constant flow rate using a syringe pump (World Precision Instrument, AL-1000). Since the viscosity of the continuous phase is high, the drops are created in air and fall into the continuous PDMS phase contained in a circular container of 2.8 cm width and 20 cm height. As the density of the PEG is higher than the density of the continuous phase, the drops sediment one by one through a height of PDMS of 10 cm before hitting the pile of the already created emulsion about $100-300$ s after generation.

\subsubsection{X-Ray tomography lab setup}\label{XRaytomo}

The tomographic device was composed of a micro-focus 150 kV Hamamatsu X-ray source with tungsten target. The sample was mounted on a precision rotation stage from Huber Germany (one circle goniometer 408) synchronised with the recording software, providing a stack of images (projections) when rotating the sample by 360$^{\circ}$. The geometrical magnification of the cone beam tomography setup is 2.5.

The sample's radioscopic projections are recorded using a flat panel detector C7942 from Hamamatsu (2240 x 2368 pixels, pixel size 50 $\micro$m). This setup results in a voxel size of (20 x 20 x 20) $\micro$m$^3$. A 60 kV acceleration voltage and a 500 $\micro$A current were found to provide the best contrast and lowest noise in the reconstructed emulsions images.

The obtained tomographs were reconstructed into a stack of horizontal slices of the 3D volume using the software \textit{Octopus}, which were then treated with the software \textit{Avizo} to obtain 3D renderings as seen in Figure \ref{fig:AngleOfRepose_tomo}b.

\subsubsection{Image Analysis}
The image analysis is performed using a C++ program following roughly the steps described in Weis S. and Schr\"oter M. \cite{Weis2017xray}. 
This allows to obtain the volume of the drops as well as their centroid position in space.
At first a bilateral filter ($\sigma_\text{g} = 40$, $\sigma_\text{p} = 4$) is used to reduce high frequency noise. 
Beam hardening artifacts are compensated by homogenizing the tomogram using the radial position of each voxel and the azimuthally averaged grey values (using 80 radial bins).
Binarisation is performed with the threshold calculated by Otsu's method. 
To treat wrongly assigned voxels due to image noise, a threedimensional median filter (filter size = 5 voxels) is used on the binary image.

The drops are labeled using an \textit{Euclidean Distance Map} (EDM) approach with an erosion depth of 35 voxels. 
After that step, all voxels have been assigned an ID which corresponds to unique drop labels or zero for air voxels (outside of the sample).
The volume of a drop is the sum of all voxels of this drop. 
We find the position of the centroid of a drop by calculating the arithmetic mean of all voxel positions of that drop.

\subsection{Results}

\subsubsection{Time evolution of the angles in the sliding experiment}

\begin{figure}[ht]
\centering
\includegraphics[width=10cm]{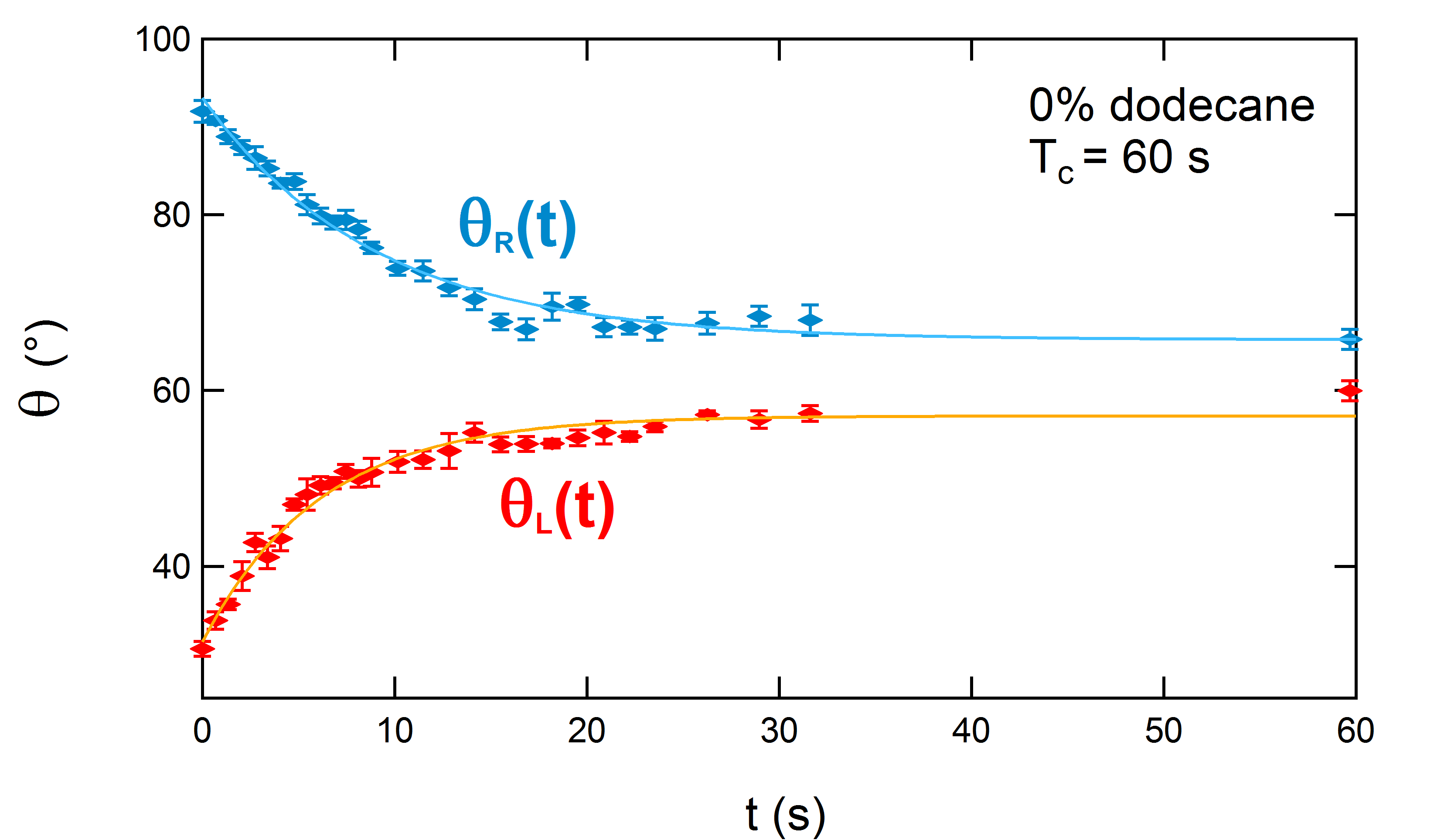}
  \caption{Evolution of the left $\theta_\text{L}$ and right $\theta_\text{R}$ angles between the top drop and the needle with time, for 0\% of dodecane and $T_\text{c}$ = 60 s. The two solid lines are fits to the experimental data.} \label{fig:Angles-fit}
\end{figure}

In Section \ref{Sec:Friction}, we analysed the relaxation towards equilibrium of the position of the drop by measuring the left $\theta_\text{L}(t)$ and right $\theta_\text{R}(t)$ angles between the drop and the needle. An example of the time evolution of $\theta_\text{L}(t)$ and $\theta_\text{R}(t)$ is given in Figure \ref{fig:Angles-fit} for 0\% of dodecane in the continuous phase and $T_\text{c}$ = 60 s. We fitted the data with exponential functions $\theta(t)=\theta_\text{f}+\theta_\text{0} \exp(-t/\tau)$, where $\theta_\text{f}$ is the value at equilibrium of the angle and $\tau$ the relaxation time.

\begin{mcitethebibliography}{86}
\providecommand*{\natexlab}[1]{#1}
\providecommand*{\mciteSetBstSublistMode}[1]{}
\providecommand*{\mciteSetBstMaxWidthForm}[2]{}
\providecommand*{\mciteBstWouldAddEndPuncttrue}
  {\def\EndOfBibitem{\unskip.}}
\providecommand*{\mciteBstWouldAddEndPunctfalse}
  {\let\EndOfBibitem\relax}
\providecommand*{\mciteSetBstMidEndSepPunct}[3]{}
\providecommand*{\mciteSetBstSublistLabelBeginEnd}[3]{}
\providecommand*{\EndOfBibitem}{}
\mciteSetBstSublistMode{f}
\mciteSetBstMaxWidthForm{subitem}
{(\emph{\alph{mcitesubitemcount}})}
\mciteSetBstSublistLabelBeginEnd{\mcitemaxwidthsubitemform\space}
{\relax}{\relax}

\bibitem[Eastoe and Dalton(2000)]{eastoe2000dynamic}
J.~Eastoe and J.~Dalton, \emph{Advances in colloid and interface science},
  2000, \textbf{85}, 103--144\relax
\mciteBstWouldAddEndPuncttrue
\mciteSetBstMidEndSepPunct{\mcitedefaultmidpunct}
{\mcitedefaultendpunct}{\mcitedefaultseppunct}\relax
\EndOfBibitem
\bibitem[Song \emph{et~al.}(2008)Song, Wang, and Makse]{Song2008}
C.~Song, P.~Wang and H.~A. Makse, \emph{Nature}, 2008, \textbf{453}, 629\relax
\mciteBstWouldAddEndPuncttrue
\mciteSetBstMidEndSepPunct{\mcitedefaultmidpunct}
{\mcitedefaultendpunct}{\mcitedefaultseppunct}\relax
\EndOfBibitem
\bibitem[Van~Hecke(2009)]{van2009jamming}
M.~Van~Hecke, \emph{Journal of Physics: Condensed Matter}, 2009, \textbf{22},
  033101\relax
\mciteBstWouldAddEndPuncttrue
\mciteSetBstMidEndSepPunct{\mcitedefaultmidpunct}
{\mcitedefaultendpunct}{\mcitedefaultseppunct}\relax
\EndOfBibitem
\bibitem[Torquato and Stillinger(2010)]{torquato2010jammed}
S.~Torquato and F.~H. Stillinger, \emph{Reviews of modern physics}, 2010,
  \textbf{82}, 2633\relax
\mciteBstWouldAddEndPuncttrue
\mciteSetBstMidEndSepPunct{\mcitedefaultmidpunct}
{\mcitedefaultendpunct}{\mcitedefaultseppunct}\relax
\EndOfBibitem
\bibitem[Andreotti \emph{et~al.}(2013)Andreotti, Forterre, and
  Pouliquen]{andreotti:13}
B.~Andreotti, Y.~Forterre and O.~Pouliquen, \emph{Granular Media}, Cambridge
  University Press, 2013\relax
\mciteBstWouldAddEndPuncttrue
\mciteSetBstMidEndSepPunct{\mcitedefaultmidpunct}
{\mcitedefaultendpunct}{\mcitedefaultseppunct}\relax
\EndOfBibitem
\bibitem[Behringer(2015)]{behringer2015jamming}
R.~P. Behringer, \emph{Comptes Rendus Physique}, 2015, \textbf{16},
  10--25\relax
\mciteBstWouldAddEndPuncttrue
\mciteSetBstMidEndSepPunct{\mcitedefaultmidpunct}
{\mcitedefaultendpunct}{\mcitedefaultseppunct}\relax
\EndOfBibitem
\bibitem[Bi \emph{et~al.}(2015)Bi, Henkes, Daniels, and
  Chakraborty]{bi2015statistical}
D.~Bi, S.~Henkes, K.~E. Daniels and B.~Chakraborty, \emph{Annu. Rev. Condens.
  Matter Phys.}, 2015, \textbf{6}, 63--83\relax
\mciteBstWouldAddEndPuncttrue
\mciteSetBstMidEndSepPunct{\mcitedefaultmidpunct}
{\mcitedefaultendpunct}{\mcitedefaultseppunct}\relax
\EndOfBibitem
\bibitem[Jaeger(2015)]{jaeger:15}
H.~M. Jaeger, \emph{Soft Matter}, 2015, \textbf{11}, 12--27\relax
\mciteBstWouldAddEndPuncttrue
\mciteSetBstMidEndSepPunct{\mcitedefaultmidpunct}
{\mcitedefaultendpunct}{\mcitedefaultseppunct}\relax
\EndOfBibitem
\bibitem[Baule \emph{et~al.}(2016)Baule, Morone, Herrmann, and
  Makse]{baule2016edwards}
A.~Baule, F.~Morone, H.~J. Herrmann and H.~A. Makse, \emph{arXiv preprint
  arXiv:1602.04369}, 2016\relax
\mciteBstWouldAddEndPuncttrue
\mciteSetBstMidEndSepPunct{\mcitedefaultmidpunct}
{\mcitedefaultendpunct}{\mcitedefaultseppunct}\relax
\EndOfBibitem
\bibitem[Luding(2016)]{luding2016granular}
S.~Luding, \emph{Nature physics}, 2016, \textbf{12}, 531--532\relax
\mciteBstWouldAddEndPuncttrue
\mciteSetBstMidEndSepPunct{\mcitedefaultmidpunct}
{\mcitedefaultendpunct}{\mcitedefaultseppunct}\relax
\EndOfBibitem
\bibitem[Weaire \emph{et~al.}(2007)Weaire, Langlois, Saadatfar, and
  Hutzler]{weaire2007foam}
D.~Weaire, V.~Langlois, M.~Saadatfar and S.~Hutzler, in \emph{Granular and
  Complex Materials}, 2007, pp. 1--26\relax
\mciteBstWouldAddEndPuncttrue
\mciteSetBstMidEndSepPunct{\mcitedefaultmidpunct}
{\mcitedefaultendpunct}{\mcitedefaultseppunct}\relax
\EndOfBibitem
\bibitem[Katgert \emph{et~al.}(2013)Katgert, Tighe, and van
  Hecke]{katgert2013jamming}
G.~Katgert, B.~P. Tighe and M.~van Hecke, \emph{Soft Matter}, 2013, \textbf{9},
  9739--9746\relax
\mciteBstWouldAddEndPuncttrue
\mciteSetBstMidEndSepPunct{\mcitedefaultmidpunct}
{\mcitedefaultendpunct}{\mcitedefaultseppunct}\relax
\EndOfBibitem
\bibitem[Schr\"oter(2017)]{schroeter:17}
M.~Schr\"oter, \emph{EPJ Web of Conferences}, 2017, \textbf{140}, 01008\relax
\mciteBstWouldAddEndPuncttrue
\mciteSetBstMidEndSepPunct{\mcitedefaultmidpunct}
{\mcitedefaultendpunct}{\mcitedefaultseppunct}\relax
\EndOfBibitem
\bibitem[Onoda and Liniger(1990)]{onoda:90}
G.~Y. Onoda and E.~G. Liniger, \emph{Phys. Rev. Lett.}, 1990, \textbf{64},
  2727--2730\relax
\mciteBstWouldAddEndPuncttrue
\mciteSetBstMidEndSepPunct{\mcitedefaultmidpunct}
{\mcitedefaultendpunct}{\mcitedefaultseppunct}\relax
\EndOfBibitem
\bibitem[Jerkins \emph{et~al.}(2008)Jerkins, Schr{\"o}ter, Swinney, Senden,
  Saadatfar, and Aste]{Jerkins2008}
M.~Jerkins, M.~Schr{\"o}ter, H.~L. Swinney, T.~J. Senden, M.~Saadatfar and
  T.~Aste, \emph{Physical review letters}, 2008, \textbf{101}, 018301\relax
\mciteBstWouldAddEndPuncttrue
\mciteSetBstMidEndSepPunct{\mcitedefaultmidpunct}
{\mcitedefaultendpunct}{\mcitedefaultseppunct}\relax
\EndOfBibitem
\bibitem[Farrell \emph{et~al.}(2010)Farrell, Martini, and Menon]{farrell:10}
G.~R. Farrell, K.~M. Martini and N.~Menon, \emph{Soft Matter}, 2010,
  \textbf{6}, 2925--2930\relax
\mciteBstWouldAddEndPuncttrue
\mciteSetBstMidEndSepPunct{\mcitedefaultmidpunct}
{\mcitedefaultendpunct}{\mcitedefaultseppunct}\relax
\EndOfBibitem
\bibitem[Silbert(2010)]{silbert:10}
L.~E. Silbert, \emph{Soft Matter}, 2010, \textbf{6}, 2918--2924\relax
\mciteBstWouldAddEndPuncttrue
\mciteSetBstMidEndSepPunct{\mcitedefaultmidpunct}
{\mcitedefaultendpunct}{\mcitedefaultseppunct}\relax
\EndOfBibitem
\bibitem[Scott and Kilgour(1969)]{Scott1969}
G.~Scott and D.~Kilgour, \emph{Journal of Physics D: Applied Physics}, 1969,
  \textbf{2}, 863\relax
\mciteBstWouldAddEndPuncttrue
\mciteSetBstMidEndSepPunct{\mcitedefaultmidpunct}
{\mcitedefaultendpunct}{\mcitedefaultseppunct}\relax
\EndOfBibitem
\bibitem[Anikeenko and Medvedev(2007)]{anikeenko_polytetrahedral_2007}
A.~V. Anikeenko and N.~N. Medvedev, \emph{Phys. Rev. Lett.}, 2007, \textbf{98},
  235504\relax
\mciteBstWouldAddEndPuncttrue
\mciteSetBstMidEndSepPunct{\mcitedefaultmidpunct}
{\mcitedefaultendpunct}{\mcitedefaultseppunct}\relax
\EndOfBibitem
\bibitem[Torquato \emph{et~al.}(2000)Torquato, Truskett, and
  Debenedetti]{torquato:00}
S.~Torquato, T.~M. Truskett and P.~G. Debenedetti, \emph{Phys. Rev. Lett.},
  2000, \textbf{84}, 2064\relax
\mciteBstWouldAddEndPuncttrue
\mciteSetBstMidEndSepPunct{\mcitedefaultmidpunct}
{\mcitedefaultendpunct}{\mcitedefaultseppunct}\relax
\EndOfBibitem
\bibitem[Jin and Makse(2010)]{jin_first-order_2010}
Y.~Jin and H.~A. Makse, \emph{Physica A}, 2010, \textbf{389}, 5362--5379\relax
\mciteBstWouldAddEndPuncttrue
\mciteSetBstMidEndSepPunct{\mcitedefaultmidpunct}
{\mcitedefaultendpunct}{\mcitedefaultseppunct}\relax
\EndOfBibitem
\bibitem[Kapfer \emph{et~al.}(2012)Kapfer, Mickel, Mecke, and
  Schr\"oder-Turk]{kapfer_jammed_2012}
S.~C. Kapfer, W.~Mickel, K.~Mecke and G.~E. Schr\"oder-Turk, \emph{Phys. Rev.
  E}, 2012, \textbf{85}, 030301\relax
\mciteBstWouldAddEndPuncttrue
\mciteSetBstMidEndSepPunct{\mcitedefaultmidpunct}
{\mcitedefaultendpunct}{\mcitedefaultseppunct}\relax
\EndOfBibitem
\bibitem[Francois \emph{et~al.}(2013)Francois, Saadatfar, Cruikshank, and
  Sheppard]{francois:13}
N.~Francois, M.~Saadatfar, R.~Cruikshank and A.~Sheppard, \emph{Phys. Rev.
  Lett.}, 2013, \textbf{111}, 148001\relax
\mciteBstWouldAddEndPuncttrue
\mciteSetBstMidEndSepPunct{\mcitedefaultmidpunct}
{\mcitedefaultendpunct}{\mcitedefaultseppunct}\relax
\EndOfBibitem
\bibitem[Baranau and Tallarek(2014)]{baranau_random_close_2014}
V.~Baranau and U.~Tallarek, \emph{Soft Matter}, 2014, \textbf{10},
  3826--3841\relax
\mciteBstWouldAddEndPuncttrue
\mciteSetBstMidEndSepPunct{\mcitedefaultmidpunct}
{\mcitedefaultendpunct}{\mcitedefaultseppunct}\relax
\EndOfBibitem
\bibitem[Rietz \emph{et~al.}(2018)Rietz, Radin, Swinney, and
  Schr\"oter]{rietz:18}
F.~Rietz, C.~Radin, H.~L. Swinney and M.~Schr\"oter, \emph{Physical Review
  Letters}, 2018, \textbf{120}, 055701\relax
\mciteBstWouldAddEndPuncttrue
\mciteSetBstMidEndSepPunct{\mcitedefaultmidpunct}
{\mcitedefaultendpunct}{\mcitedefaultseppunct}\relax
\EndOfBibitem
\bibitem[Weaire and Hutzler(1999)]{weaire1999physics}
D.~Weaire and S.~Hutzler, \emph{New York}, 1999\relax
\mciteBstWouldAddEndPuncttrue
\mciteSetBstMidEndSepPunct{\mcitedefaultmidpunct}
{\mcitedefaultendpunct}{\mcitedefaultseppunct}\relax
\EndOfBibitem
\bibitem[Cantat \emph{et~al.}(2013)Cantat, Cohen-Addad, Elias, Graner,
  H{\"o}hler, Pitois, Rouyer, and Saint-Jalmes]{cantat2013foams}
I.~Cantat, S.~Cohen-Addad, F.~Elias, F.~Graner, R.~H{\"o}hler, O.~Pitois,
  F.~Rouyer and A.~Saint-Jalmes, \emph{Foams: structure and dynamics}, OUP
  Oxford, 2013\relax
\mciteBstWouldAddEndPuncttrue
\mciteSetBstMidEndSepPunct{\mcitedefaultmidpunct}
{\mcitedefaultendpunct}{\mcitedefaultseppunct}\relax
\EndOfBibitem
\bibitem[Drenckhan and Hutzler(2015)]{drenckhan2015structure}
W.~Drenckhan and S.~Hutzler, \emph{Advances in colloid and interface science},
  2015, \textbf{224}, 1--16\relax
\mciteBstWouldAddEndPuncttrue
\mciteSetBstMidEndSepPunct{\mcitedefaultmidpunct}
{\mcitedefaultendpunct}{\mcitedefaultseppunct}\relax
\EndOfBibitem
\bibitem[H{\"o}hler and Cohen-Addad(2017)]{hohler2017many}
R.~H{\"o}hler and S.~Cohen-Addad, \emph{Soft matter}, 2017, \textbf{13},
  1371--1383\relax
\mciteBstWouldAddEndPuncttrue
\mciteSetBstMidEndSepPunct{\mcitedefaultmidpunct}
{\mcitedefaultendpunct}{\mcitedefaultseppunct}\relax
\EndOfBibitem
\bibitem[Weaire \emph{et~al.}(2017)Weaire, H{\"o}hler, and
  Hutzler]{weaire2017bubble}
D.~Weaire, R.~H{\"o}hler and S.~Hutzler, \emph{Advances in colloid and
  interface science}, 2017, \textbf{247}, 491--495\relax
\mciteBstWouldAddEndPuncttrue
\mciteSetBstMidEndSepPunct{\mcitedefaultmidpunct}
{\mcitedefaultendpunct}{\mcitedefaultseppunct}\relax
\EndOfBibitem
\bibitem[Rio \emph{et~al.}(2014)Rio, Drenckhan, Salonen, and
  Langevin]{rio2014unusually}
E.~Rio, W.~Drenckhan, A.~Salonen and D.~Langevin, \emph{Advances in colloid and
  interface science}, 2014, \textbf{205}, 74--86\relax
\mciteBstWouldAddEndPuncttrue
\mciteSetBstMidEndSepPunct{\mcitedefaultmidpunct}
{\mcitedefaultendpunct}{\mcitedefaultseppunct}\relax
\EndOfBibitem
\bibitem[Salonen \emph{et~al.}(2014)Salonen, Drenckhan, and
  Rio]{salonen2014interfacial}
A.~Salonen, W.~Drenckhan and E.~Rio, \emph{Soft matter}, 2014, \textbf{10},
  6870--6872\relax
\mciteBstWouldAddEndPuncttrue
\mciteSetBstMidEndSepPunct{\mcitedefaultmidpunct}
{\mcitedefaultendpunct}{\mcitedefaultseppunct}\relax
\EndOfBibitem
\bibitem[Heim \emph{et~al.}(2015)Heim, Vinod-Kumar, Garc{\'\i}a-Moreno, Rack,
  and Banhart]{heim2015stabilisation}
K.~Heim, G.~Vinod-Kumar, F.~Garc{\'\i}a-Moreno, A.~Rack and J.~Banhart,
  \emph{Acta Materialia}, 2015, \textbf{99}, 313--324\relax
\mciteBstWouldAddEndPuncttrue
\mciteSetBstMidEndSepPunct{\mcitedefaultmidpunct}
{\mcitedefaultendpunct}{\mcitedefaultseppunct}\relax
\EndOfBibitem
\bibitem[Heim \emph{et~al.}(2014)Heim, Garc{\'\i}a-Moreno, Kumar, Rack, and
  Banhart]{heim2014rupture}
K.~Heim, F.~Garc{\'\i}a-Moreno, G.~V. Kumar, A.~Rack and J.~Banhart, \emph{Soft
  matter}, 2014, \textbf{10}, 4711--4716\relax
\mciteBstWouldAddEndPuncttrue
\mciteSetBstMidEndSepPunct{\mcitedefaultmidpunct}
{\mcitedefaultendpunct}{\mcitedefaultseppunct}\relax
\EndOfBibitem
\bibitem[Giustiniani \emph{et~al.}(2016)Giustiniani, Gu{\'e}gan, Marchand,
  Poulard, and Drenckhan]{Giustiniani2016}
A.~Giustiniani, P.~Gu{\'e}gan, M.~Marchand, C.~Poulard and W.~Drenckhan,
  \emph{Macromolecular Rapid Communications}, 2016, \textbf{37},
  1527--1532\relax
\mciteBstWouldAddEndPuncttrue
\mciteSetBstMidEndSepPunct{\mcitedefaultmidpunct}
{\mcitedefaultendpunct}{\mcitedefaultseppunct}\relax
\EndOfBibitem
\bibitem[Feng \emph{et~al.}(2013)Feng, Pontani, Dreyfus, Chaikin, and
  Brujic]{Feng2013}
L.~Feng, L.-L. Pontani, R.~Dreyfus, P.~Chaikin and J.~Brujic, \emph{Soft
  Matter}, 2013, \textbf{9}, 9816--9823\relax
\mciteBstWouldAddEndPuncttrue
\mciteSetBstMidEndSepPunct{\mcitedefaultmidpunct}
{\mcitedefaultendpunct}{\mcitedefaultseppunct}\relax
\EndOfBibitem
\bibitem[Pontani \emph{et~al.}(2012)Pontani, Jorjadze, Viasnoff, and
  Brujic]{Pontani2012}
L.-L. Pontani, I.~Jorjadze, V.~Viasnoff and J.~Brujic, \emph{Proceedings of the
  National Academy of Sciences}, 2012, \textbf{109}, 9839--9844\relax
\mciteBstWouldAddEndPuncttrue
\mciteSetBstMidEndSepPunct{\mcitedefaultmidpunct}
{\mcitedefaultendpunct}{\mcitedefaultseppunct}\relax
\EndOfBibitem
\bibitem[Pontani \emph{et~al.}(2013)Pontani, Haase, Raczkowska, and
  Brujic]{Pontani2013}
L.-L. Pontani, M.~F. Haase, I.~Raczkowska and J.~Brujic, \emph{Soft Matter},
  2013, \textbf{9}, 7150--7157\relax
\mciteBstWouldAddEndPuncttrue
\mciteSetBstMidEndSepPunct{\mcitedefaultmidpunct}
{\mcitedefaultendpunct}{\mcitedefaultseppunct}\relax
\EndOfBibitem
\bibitem[Pontani \emph{et~al.}(2016)Pontani, Jorjadze, and Brujic]{Pontani2016}
L.-L. Pontani, I.~Jorjadze and J.~Brujic, \emph{Biophysical journal}, 2016,
  \textbf{110}, 391--399\relax
\mciteBstWouldAddEndPuncttrue
\mciteSetBstMidEndSepPunct{\mcitedefaultmidpunct}
{\mcitedefaultendpunct}{\mcitedefaultseppunct}\relax
\EndOfBibitem
\bibitem[Jorjadze \emph{et~al.}(2011)Jorjadze, Pontani, Newhall, and
  Bruji{\'c}]{Jorjadze2011}
I.~Jorjadze, L.-L. Pontani, K.~A. Newhall and J.~Bruji{\'c}, \emph{Proceedings
  of the National Academy of Sciences}, 2011, \textbf{108}, 4286--4291\relax
\mciteBstWouldAddEndPuncttrue
\mciteSetBstMidEndSepPunct{\mcitedefaultmidpunct}
{\mcitedefaultendpunct}{\mcitedefaultseppunct}\relax
\EndOfBibitem
\bibitem[Hadorn \emph{et~al.}(2012)Hadorn, Boenzli, S{\o}rensen, Fellermann,
  Hotz, and Hanczyc]{Hadorn2012}
M.~Hadorn, E.~Boenzli, K.~T. S{\o}rensen, H.~Fellermann, P.~E. Hotz and M.~M.
  Hanczyc, \emph{Proceedings of the National Academy of Sciences}, 2012,
  \textbf{109}, 20320--20325\relax
\mciteBstWouldAddEndPuncttrue
\mciteSetBstMidEndSepPunct{\mcitedefaultmidpunct}
{\mcitedefaultendpunct}{\mcitedefaultseppunct}\relax
\EndOfBibitem
\bibitem[Papanikolaou \emph{et~al.}(2013)Papanikolaou, O'Hern, and
  Shattuck]{papanikolaou:13}
S.~Papanikolaou, C.~S. O'Hern and M.~D. Shattuck, \emph{Physical Review
  Letters}, 2013, \textbf{110}, 198002\relax
\mciteBstWouldAddEndPuncttrue
\mciteSetBstMidEndSepPunct{\mcitedefaultmidpunct}
{\mcitedefaultendpunct}{\mcitedefaultseppunct}\relax
\EndOfBibitem
\bibitem[Liu \emph{et~al.}(2015)Liu, Li, Baule, and Makse]{liu:15}
W.~Liu, S.~Li, A.~Baule and H.~A. Makse, \emph{Soft Matter}, 2015, \textbf{11},
  6492--6498\relax
\mciteBstWouldAddEndPuncttrue
\mciteSetBstMidEndSepPunct{\mcitedefaultmidpunct}
{\mcitedefaultendpunct}{\mcitedefaultseppunct}\relax
\EndOfBibitem
\bibitem[Chen \emph{et~al.}(2016)Chen, Li, Liu, and A.Makse]{chen:16}
S.~Chen, S.~Li, W.~Liu and H.~A.Makse, \emph{Soft Matter}, 2016, \textbf{12},
  1836--1846\relax
\mciteBstWouldAddEndPuncttrue
\mciteSetBstMidEndSepPunct{\mcitedefaultmidpunct}
{\mcitedefaultendpunct}{\mcitedefaultseppunct}\relax
\EndOfBibitem
\bibitem[Mughal \emph{et~al.}(2012)Mughal, Chan, Weaire, and
  Hutzler]{mughal2012dense}
A.~Mughal, H.~Chan, D.~Weaire and S.~Hutzler, \emph{Physical Review E}, 2012,
  \textbf{85}, 051305\relax
\mciteBstWouldAddEndPuncttrue
\mciteSetBstMidEndSepPunct{\mcitedefaultmidpunct}
{\mcitedefaultendpunct}{\mcitedefaultseppunct}\relax
\EndOfBibitem
\bibitem[Scheel \emph{et~al.}(2008)Scheel, Seemann, Brinkmann, Di~Michiel,
  Sheppard, Breidenbach, and Herminghaus]{scheel:08}
M.~Scheel, R.~Seemann, M.~Brinkmann, M.~Di~Michiel, A.~Sheppard, B.~Breidenbach
  and S.~Herminghaus, \emph{Nature Materials}, 2008, \textbf{7}, 189--193\relax
\mciteBstWouldAddEndPuncttrue
\mciteSetBstMidEndSepPunct{\mcitedefaultmidpunct}
{\mcitedefaultendpunct}{\mcitedefaultseppunct}\relax
\EndOfBibitem
\bibitem[Lois \emph{et~al.}(2008)Lois, Blawzdziewicz, and O'Hern]{lois:08}
G.~Lois, J.~Blawzdziewicz and C.~S. O'Hern, \emph{Physical Review Letters},
  2008, \textbf{100}, 028001\relax
\mciteBstWouldAddEndPuncttrue
\mciteSetBstMidEndSepPunct{\mcitedefaultmidpunct}
{\mcitedefaultendpunct}{\mcitedefaultseppunct}\relax
\EndOfBibitem
\bibitem[Butt and Kappl(2009)]{butt:09}
H.-J. Butt and M.~Kappl, \emph{Advances in Colloid and Interface Science},
  2009, \textbf{146}, 48--60\relax
\mciteBstWouldAddEndPuncttrue
\mciteSetBstMidEndSepPunct{\mcitedefaultmidpunct}
{\mcitedefaultendpunct}{\mcitedefaultseppunct}\relax
\EndOfBibitem
\bibitem[G\"ogelein \emph{et~al.}(2010)G\"ogelein, Brinkmann, Schr\"oter, and
  Herminghaus]{gogelein:10}
C.~G\"ogelein, M.~Brinkmann, M.~Schr\"oter and S.~Herminghaus, \emph{Langmuir},
  2010, \textbf{26}, 17184--17189\relax
\mciteBstWouldAddEndPuncttrue
\mciteSetBstMidEndSepPunct{\mcitedefaultmidpunct}
{\mcitedefaultendpunct}{\mcitedefaultseppunct}\relax
\EndOfBibitem
\bibitem[Koos and Willenbacher(2012)]{koos:12}
E.~Koos and N.~Willenbacher, \emph{Soft Matter}, 2012, \textbf{8},
  3988--3994\relax
\mciteBstWouldAddEndPuncttrue
\mciteSetBstMidEndSepPunct{\mcitedefaultmidpunct}
{\mcitedefaultendpunct}{\mcitedefaultseppunct}\relax
\EndOfBibitem
\bibitem[Rieser \emph{et~al.}(2015)Rieser, Arratia, Yodh, Gollub, and
  Durian]{rieser:15}
J.~M. Rieser, P.~E. Arratia, A.~G. Yodh, J.~P. Gollub and D.~J. Durian,
  \emph{Langmuir}, 2015, \textbf{31}, 2421--2429\relax
\mciteBstWouldAddEndPuncttrue
\mciteSetBstMidEndSepPunct{\mcitedefaultmidpunct}
{\mcitedefaultendpunct}{\mcitedefaultseppunct}\relax
\EndOfBibitem
\bibitem[Hemmerle \emph{et~al.}(2016)Hemmerle, Schr\"oter, and
  Goehring]{hemmerle:16}
A.~Hemmerle, M.~Schr\"oter and L.~Goehring, \emph{Scientific Reports}, 2016,
  \textbf{6}, 35650\relax
\mciteBstWouldAddEndPuncttrue
\mciteSetBstMidEndSepPunct{\mcitedefaultmidpunct}
{\mcitedefaultendpunct}{\mcitedefaultseppunct}\relax
\EndOfBibitem
\bibitem[Pohlman \emph{et~al.}(2006)Pohlman, Severson, Ottino, and
  Lueptow]{pohlman:06}
N.~A. Pohlman, B.~L. Severson, J.~M. Ottino and R.~M. Lueptow, \emph{Physical
  Review E (Statistical, Nonlinear, and Soft Matter Physics)}, 2006,
  \textbf{73}, 031304--9\relax
\mciteBstWouldAddEndPuncttrue
\mciteSetBstMidEndSepPunct{\mcitedefaultmidpunct}
{\mcitedefaultendpunct}{\mcitedefaultseppunct}\relax
\EndOfBibitem
\bibitem[Utermann \emph{et~al.}(2011)Utermann, Aurin, Benderoth, Fischer, and
  Schr\"oter]{utermann:11}
S.~Utermann, P.~Aurin, M.~Benderoth, C.~Fischer and M.~Schr\"oter, \emph{Phys.
  Rev. E}, 2011, \textbf{84}, 031306\relax
\mciteBstWouldAddEndPuncttrue
\mciteSetBstMidEndSepPunct{\mcitedefaultmidpunct}
{\mcitedefaultendpunct}{\mcitedefaultseppunct}\relax
\EndOfBibitem
\bibitem[Back(2011)]{back:11}
R.~Back, \emph{Granular Matter}, 2011, \textbf{13}, 723--729\relax
\mciteBstWouldAddEndPuncttrue
\mciteSetBstMidEndSepPunct{\mcitedefaultmidpunct}
{\mcitedefaultendpunct}{\mcitedefaultseppunct}\relax
\EndOfBibitem
\bibitem[Sheng \emph{et~al.}(2016)Sheng, Chang, and Hsiau]{sheng:16}
L.-T. Sheng, W.-C. Chang and S.-S. Hsiau, \emph{Physical Review E}, 2016,
  \textbf{94}, 012903\relax
\mciteBstWouldAddEndPuncttrue
\mciteSetBstMidEndSepPunct{\mcitedefaultmidpunct}
{\mcitedefaultendpunct}{\mcitedefaultseppunct}\relax
\EndOfBibitem
\bibitem[Gillemot \emph{et~al.}(2017)Gillemot, Somfai, and
  B\"orzs\"onyi]{gillemot:17}
K.~A. Gillemot, E.~Somfai and T.~B\"orzs\"onyi, \emph{Soft Matter}, 2017,
  \textbf{13}, 415--420\relax
\mciteBstWouldAddEndPuncttrue
\mciteSetBstMidEndSepPunct{\mcitedefaultmidpunct}
{\mcitedefaultendpunct}{\mcitedefaultseppunct}\relax
\EndOfBibitem
\bibitem[Maeda \emph{et~al.}(2002)Maeda, Chen, Tirrell, and
  Israelachvili]{maeda2002adhesion}
N.~Maeda, N.~Chen, M.~Tirrell and J.~N. Israelachvili, \emph{Science}, 2002,
  \textbf{297}, 379--382\relax
\mciteBstWouldAddEndPuncttrue
\mciteSetBstMidEndSepPunct{\mcitedefaultmidpunct}
{\mcitedefaultendpunct}{\mcitedefaultseppunct}\relax
\EndOfBibitem
\bibitem[Myshkin \emph{et~al.}(2006)Myshkin, Petrokovets, and
  Kovalev]{Myshkin2006}
N.~Myshkin, M.~Petrokovets and A.~Kovalev, \emph{Tribology International},
  2006, \textbf{38}, 910--921\relax
\mciteBstWouldAddEndPuncttrue
\mciteSetBstMidEndSepPunct{\mcitedefaultmidpunct}
{\mcitedefaultendpunct}{\mcitedefaultseppunct}\relax
\EndOfBibitem
\bibitem[Lespiat \emph{et~al.}(2011)Lespiat, Cohen-Addad, and
  H{\"o}hler]{lespiat2011jamming}
R.~Lespiat, S.~Cohen-Addad and R.~H{\"o}hler, \emph{Physical review letters},
  2011, \textbf{106}, 148302\relax
\mciteBstWouldAddEndPuncttrue
\mciteSetBstMidEndSepPunct{\mcitedefaultmidpunct}
{\mcitedefaultendpunct}{\mcitedefaultseppunct}\relax
\EndOfBibitem
\bibitem[Peyneau and Roux(2008)]{peyneau2008frictionless}
P.-E. Peyneau and J.-N. Roux, \emph{Physical review E}, 2008, \textbf{78},
  011307\relax
\mciteBstWouldAddEndPuncttrue
\mciteSetBstMidEndSepPunct{\mcitedefaultmidpunct}
{\mcitedefaultendpunct}{\mcitedefaultseppunct}\relax
\EndOfBibitem
\bibitem[Taboada \emph{et~al.}(2006)Taboada, Estrada, and
  Radjai]{taboada2006additive}
A.~Taboada, N.~Estrada and F.~Radjai, \emph{Physical review letters}, 2006,
  \textbf{97}, 098302\relax
\mciteBstWouldAddEndPuncttrue
\mciteSetBstMidEndSepPunct{\mcitedefaultmidpunct}
{\mcitedefaultendpunct}{\mcitedefaultseppunct}\relax
\EndOfBibitem
\bibitem[Weis and Schr\"oter(2017)]{Weis2017xray}
S.~Weis and M.~Schr\"oter, \emph{Review of Scientific Instruments}, 2017,
  \textbf{88}, 051809\relax
\mciteBstWouldAddEndPuncttrue
\mciteSetBstMidEndSepPunct{\mcitedefaultmidpunct}
{\mcitedefaultendpunct}{\mcitedefaultseppunct}\relax
\EndOfBibitem
\bibitem[Schaller \emph{et~al.}(2013)Schaller, Kapfer, Evans, Hoffmann, Aste,
  Saadatfar, Mecke, Delaney, and Schr\"oder-Turk]{Schaller2013SetVoronoi}
F.~M. Schaller, S.~C. Kapfer, M.~E. Evans, M.~J. Hoffmann, T.~Aste,
  M.~Saadatfar, K.~Mecke, G.~W. Delaney and G.~E. Schr\"oder-Turk,
  \emph{Philosophical Magazine}, 2013, \textbf{93}, 3993 -- 4017\relax
\mciteBstWouldAddEndPuncttrue
\mciteSetBstMidEndSepPunct{\mcitedefaultmidpunct}
{\mcitedefaultendpunct}{\mcitedefaultseppunct}\relax
\EndOfBibitem
\bibitem[Weis \emph{et~al.}(2017)Weis, Sch\"onh\"ofer, Schaller, Schr\"oter,
  and Schr\"oder-Turk]{weis:17}
S.~Weis, P.~W.~A. Sch\"onh\"ofer, F.~M. Schaller, M.~Schr\"oter and G.~E.
  Schr\"oder-Turk, \emph{EPJ Web of Conferences}, 2017, \textbf{140},
  06007\relax
\mciteBstWouldAddEndPuncttrue
\mciteSetBstMidEndSepPunct{\mcitedefaultmidpunct}
{\mcitedefaultendpunct}{\mcitedefaultseppunct}\relax
\EndOfBibitem
\bibitem[Maestro \emph{et~al.}(2013)Maestro, Drenckhan, Rio, and
  H{\"o}hler]{Maestro2013}
A.~Maestro, W.~Drenckhan, E.~Rio and R.~H{\"o}hler, \emph{Soft Matter}, 2013,
  \textbf{9}, 2531--2540\relax
\mciteBstWouldAddEndPuncttrue
\mciteSetBstMidEndSepPunct{\mcitedefaultmidpunct}
{\mcitedefaultendpunct}{\mcitedefaultseppunct}\relax
\EndOfBibitem
\bibitem[H{\"o}hler \emph{et~al.}(2008)H{\"o}hler, Yip Cheung~Sang, Lorenceau,
  and Cohen-Addad]{Hohler2008}
R.~H{\"o}hler, Y.~Yip Cheung~Sang, E.~Lorenceau and S.~Cohen-Addad,
  \emph{Langmuir}, 2008, \textbf{24}, 418--425\relax
\mciteBstWouldAddEndPuncttrue
\mciteSetBstMidEndSepPunct{\mcitedefaultmidpunct}
{\mcitedefaultendpunct}{\mcitedefaultseppunct}\relax
\EndOfBibitem
\bibitem[Giustiniani(2017)]{giustiniani_thesis}
A.~Giustiniani, \emph{Theses}, {Universit{\'e} Paris-Saclay}, 2017\relax
\mciteBstWouldAddEndPuncttrue
\mciteSetBstMidEndSepPunct{\mcitedefaultmidpunct}
{\mcitedefaultendpunct}{\mcitedefaultseppunct}\relax
\EndOfBibitem
\bibitem[Arditty \emph{et~al.}(2003)Arditty, Whitby, Binks, Schmitt, and
  Leal-Calderon]{arditty2003some}
S.~Arditty, C.~P. Whitby, B.~Binks, V.~Schmitt and F.~Leal-Calderon, \emph{The
  European Physical Journal E}, 2003, \textbf{11}, 273--281\relax
\mciteBstWouldAddEndPuncttrue
\mciteSetBstMidEndSepPunct{\mcitedefaultmidpunct}
{\mcitedefaultendpunct}{\mcitedefaultseppunct}\relax
\EndOfBibitem
\bibitem[Janssen(1895)]{janssen:95}
H.~Janssen, \emph{Zeitschrift des Vereines deutscher Ingenieure}, 1895,
  \textbf{39}, 1045--1049\relax
\mciteBstWouldAddEndPuncttrue
\mciteSetBstMidEndSepPunct{\mcitedefaultmidpunct}
{\mcitedefaultendpunct}{\mcitedefaultseppunct}\relax
\EndOfBibitem
\bibitem[Sperl(2006)]{sperl:06}
M.~Sperl, \emph{Granular Matter}, 2006, \textbf{8}, 59--65\relax
\mciteBstWouldAddEndPuncttrue
\mciteSetBstMidEndSepPunct{\mcitedefaultmidpunct}
{\mcitedefaultendpunct}{\mcitedefaultseppunct}\relax
\EndOfBibitem
\bibitem[Vanel and Cl\'ement(1999)]{vanel:99}
L.~Vanel and E.~Cl\'ement, \emph{The European Physical Journal B}, 1999,
  \textbf{11}, 525--533\relax
\mciteBstWouldAddEndPuncttrue
\mciteSetBstMidEndSepPunct{\mcitedefaultmidpunct}
{\mcitedefaultendpunct}{\mcitedefaultseppunct}\relax
\EndOfBibitem
\bibitem[Bertho \emph{et~al.}(2003)Bertho, Giorgiutti-Dauphin{\'e}, and
  Hulin]{Bertho2003}
Y.~Bertho, F.~Giorgiutti-Dauphin{\'e} and J.-P. Hulin, \emph{Physical review
  letters}, 2003, \textbf{90}, 144301\relax
\mciteBstWouldAddEndPuncttrue
\mciteSetBstMidEndSepPunct{\mcitedefaultmidpunct}
{\mcitedefaultendpunct}{\mcitedefaultseppunct}\relax
\EndOfBibitem
\bibitem[Wambaugh \emph{et~al.}(2010)Wambaugh, Hartley, and
  Behringer]{wambaugh:10}
J.~Wambaugh, R.~Hartley and R.~Behringer, \emph{The European Physical Journal
  E}, 2010, \textbf{32}, 135--145\relax
\mciteBstWouldAddEndPuncttrue
\mciteSetBstMidEndSepPunct{\mcitedefaultmidpunct}
{\mcitedefaultendpunct}{\mcitedefaultseppunct}\relax
\EndOfBibitem
\bibitem[Creton(2003)]{creton2003pressure}
C.~Creton, \emph{MRS bulletin}, 2003, \textbf{28}, 434--439\relax
\mciteBstWouldAddEndPuncttrue
\mciteSetBstMidEndSepPunct{\mcitedefaultmidpunct}
{\mcitedefaultendpunct}{\mcitedefaultseppunct}\relax
\EndOfBibitem
\bibitem[Liu \emph{et~al.}(2011)Liu, Yang, and Yu]{Liu2011}
P.~Liu, R.~Yang and A.~Yu, \emph{Physics of fluids}, 2011, \textbf{23},
  013304\relax
\mciteBstWouldAddEndPuncttrue
\mciteSetBstMidEndSepPunct{\mcitedefaultmidpunct}
{\mcitedefaultendpunct}{\mcitedefaultseppunct}\relax
\EndOfBibitem
\bibitem[Scales \emph{et~al.}(1998)Scales, Johnson, Healy, and
  Kapur]{scales1998shear}
P.~J. Scales, S.~B. Johnson, T.~W. Healy and P.~C. Kapur, \emph{AIChE Journal},
  1998, \textbf{44}, 538--544\relax
\mciteBstWouldAddEndPuncttrue
\mciteSetBstMidEndSepPunct{\mcitedefaultmidpunct}
{\mcitedefaultendpunct}{\mcitedefaultseppunct}\relax
\EndOfBibitem
\bibitem[Sonntag and Russel(1987)]{sonntag1987elastic}
R.~Sonntag and W.~Russel, \emph{Journal of Colloid and Interface Science},
  1987, \textbf{116}, 485--489\relax
\mciteBstWouldAddEndPuncttrue
\mciteSetBstMidEndSepPunct{\mcitedefaultmidpunct}
{\mcitedefaultendpunct}{\mcitedefaultseppunct}\relax
\EndOfBibitem
\bibitem[Zhang and Makse(2005)]{zhang2005jamming}
H.~Zhang and H.~Makse, \emph{Physical Review E}, 2005, \textbf{72},
  011301\relax
\mciteBstWouldAddEndPuncttrue
\mciteSetBstMidEndSepPunct{\mcitedefaultmidpunct}
{\mcitedefaultendpunct}{\mcitedefaultseppunct}\relax
\EndOfBibitem
\bibitem[Schr{\"o}der-Turk \emph{et~al.}(2011)Schr{\"o}der-Turk, Mickel,
  Kapfer, Klatt, Schaller, Hoffmann, Kleppmann, Armstrong, Inayat,
  Hug,\emph{et~al.}]{Schroder2011}
G.~Schr{\"o}der-Turk, W.~Mickel, S.~Kapfer, M.~Klatt, F.~Schaller, M.~Hoffmann,
  N.~Kleppmann, P.~Armstrong, A.~Inayat, D.~Hug \emph{et~al.}, \emph{Advanced
  Materials}, 2011, \textbf{23}, 2535--2553\relax
\mciteBstWouldAddEndPuncttrue
\mciteSetBstMidEndSepPunct{\mcitedefaultmidpunct}
{\mcitedefaultendpunct}{\mcitedefaultseppunct}\relax
\EndOfBibitem
\bibitem[Schr\"oder-Turk \emph{et~al.}(2010)Schr\"oder-Turk, Kapfer,
  Breidenbach, Beisbart, and Mecke]{Gerd:2009_mink}
G.~E. Schr\"oder-Turk, S.~C. Kapfer, B.~Breidenbach, C.~Beisbart and K.~Mecke,
  \emph{J.~Microsc.}, 2010, \textbf{238}, 57--74\relax
\mciteBstWouldAddEndPuncttrue
\mciteSetBstMidEndSepPunct{\mcitedefaultmidpunct}
{\mcitedefaultendpunct}{\mcitedefaultseppunct}\relax
\EndOfBibitem
\bibitem[Kapfer(2011)]{Kapfer_2011}
S.~Kapfer, \emph{PhD thesis, FAU Erlangen}, 2011\relax
\mciteBstWouldAddEndPuncttrue
\mciteSetBstMidEndSepPunct{\mcitedefaultmidpunct}
{\mcitedefaultendpunct}{\mcitedefaultseppunct}\relax
\EndOfBibitem
\bibitem[Anthony and Marone(2005)]{Anthony2005}
J.~L. Anthony and C.~Marone, \emph{Journal of Geophysical Research: Solid
  Earth}, 2005, \textbf{110}, 1--14\relax
\mciteBstWouldAddEndPuncttrue
\mciteSetBstMidEndSepPunct{\mcitedefaultmidpunct}
{\mcitedefaultendpunct}{\mcitedefaultseppunct}\relax
\EndOfBibitem
\bibitem[Brodu \emph{et~al.}(2015)Brodu, Dijksman, and Behringer]{brodu:15}
N.~Brodu, J.~A. Dijksman and R.~P. Behringer, \emph{Nature Communications},
  2015, \textbf{6}, 6361\relax
\mciteBstWouldAddEndPuncttrue
\mciteSetBstMidEndSepPunct{\mcitedefaultmidpunct}
{\mcitedefaultendpunct}{\mcitedefaultseppunct}\relax
\EndOfBibitem
\bibitem[Majmudar and Behringer(2005)]{majmudar:05}
T.~S. Majmudar and R.~P. Behringer, \emph{Nature}, 2005, \textbf{435},
  1079--1082\relax
\mciteBstWouldAddEndPuncttrue
\mciteSetBstMidEndSepPunct{\mcitedefaultmidpunct}
{\mcitedefaultendpunct}{\mcitedefaultseppunct}\relax
\EndOfBibitem
\bibitem[Daniels \emph{et~al.}(2017)Daniels, Kollmer, and Puckett]{daniels:17}
K.~E. Daniels, J.~E. Kollmer and J.~G. Puckett, \emph{Review of Scientific
  Instruments}, 2017, \textbf{88}, 051808\relax
\mciteBstWouldAddEndPuncttrue
\mciteSetBstMidEndSepPunct{\mcitedefaultmidpunct}
{\mcitedefaultendpunct}{\mcitedefaultseppunct}\relax
\EndOfBibitem
\end{mcitethebibliography}

\providecommand*{\mcitethebibliography}{\thebibliography}
\csname @ifundefined\endcsname{endmcitethebibliography}
{\let\endmcitethebibliography\endthebibliography}{}

\end{document}